\documentclass[aps, pre, twocolumn, amsmath, superscriptaddress,showkeys,showpacs]{revtex4-1}
\usepackage{amsmath, amssymb}
\usepackage{natbib}
\usepackage{dcolumn}
\usepackage{bm}
\usepackage[utf8]{inputenc}
\usepackage[T1]{fontenc}
\usepackage{mathptmx}
\usepackage{tikz}
\usepackage{color}
\usepackage{marvosym}        
\usepackage{subcaption}             
\usepackage{graphicx}
\usepackage{csquotes}
\usepackage{mathtools}
\usepackage[title]{appendix}
\usepackage{url}
\DeclareMathOperator\erf{erf}
\makeatletter
\renewcommand\@makecaption[2]{%
	\par
	\vskip\abovecaptionskip
	\begingroup
	\small\rmfamily
	\begingroup
	\samepage
	\flushing
	\let\footnote\@footnotemark@gobble
	\@make@capt@title{#1}{#2}\par
	\endgroup
	\endgroup
	\vskip\belowcaptionskip
}
\makeatother
\renewcommand\labelenumi{(\roman{enumi})}
\renewcommand\theenumi\labelenumi 
\makeatletter
\renewcommand\paragraph{\@startsection{paragraph}{4}{\z@}%
	{-2.5ex\@plus -1ex \@minus -.25ex}%
	{1.25ex \@plus .25ex}%
	{\itshape\normalsize}}
\makeatother
\usepackage{hyperref} 
\usepackage{breakurl}
\hypersetup{
	colorlinks=true,
	linkcolor=blue,
	filecolor=magenta,      
	urlcolor=cyan,
	citecolor=blue,
}
\begin{document}	
	\captionsetup[figure]{labelfont={bf},labelformat={default},labelsep=space,name={Fig.}}
	\DeclareRobustCommand\blackline{\tikz[baseline=-0.6ex]\draw[black,thick] (0,0)--(0.54,0);}
	\DeclareRobustCommand\redline{\tikz[baseline=-0.6ex]\draw[red,thick] (0,0)--(0.54,0);}
	\DeclareRobustCommand\blueline{\tikz[baseline=-0.6ex]\draw[blue,thick] (0,0)--(0.54,0);}
	\DeclareRobustCommand\greenline{\tikz[baseline=-0.6ex]\draw[green,thick] (0,0)--(0.54,0);}
	\DeclareRobustCommand\magentaline{\tikz[baseline=-0.6ex]\draw[magenta,thick] (0,0)--(0.54,0);}
	\DeclareRobustCommand\dashedred{\tikz[baseline=-0.6ex]\draw[red,thick,dashed] (0,0)--(0.54,0);}
	\DeclareRobustCommand\chainblack{\tikz[baseline=-0.6ex]\draw[black,thick,dash dot dot] (0,0)--(0.5,0);}
	\DeclareRobustCommand\chainblue{\tikz[baseline=-0.6ex]\draw[blue,thick,dash dot dot] (0,0)--(0.5,0);}
	\title{Effect of non-local grazing on dry-land vegetation dynamics}

	\author{Mrinal Kanti Pal}\email{mrinalkantipal13@gmail.com}
\author{Swarup Poria}%
\affiliation{ 
	Department of Applied Mathematics,
	University of Calcutta,
	92 APC Road, Kolkata-700009, India
}%

	\date{\today}
	\thanks{This article has been accepted for publication in \textit{Physical Review E}, link: \url{https://journals.aps.org/pre/accepted/23078R21G221d52f82055f69e43409606ae97775c}}
	\begin{abstract}
	Dry-land ecosystem has turned into a matter of grave concern, due to growing threat of land degradation and bioproductivity-loss. Self-organized vegetation patterns are a remarkable characteristic of these ecosystems; apart from being visually captivating, patterns modulate the system-response to increasing environmental stress. Empirical studies hinted that herbivory is one the key regulatory mechanism behind pattern formation and overall ecosystem functioning. However most of the mathematical models have taken a mean-field strategy to grazing; foraging has been considered to be independent of spatial distribution of vegetation. To this end, an extended version of the celebrated plant-water model due to Klausmeier, has been taken as the base here. To encompass the effect of heterogeneous vegetation distribution on foraging intensity and subsequent impact on entire ecosystem, grazing is considered here to depend on spatially weighted average vegetation density, instead of density at a particular point. Moreover, varying influence of vegetation at any location over gazing elsewhere, is incorporated by choosing suitable averaging function. A comprehensive analysis demonstrates that inclusion of spatial non-locality, alters the understanding of system dynamics significantly. The grazing ecosystem is found to be more resilient to increasing aridity than it was anticipated to be in earlier studies on non-local grazing. The system-response to rising environmental pressure is also observed to vary depending on the grazer. Obtained results also suggest possibility of multi-stability due the history-dependence of system-response. Overall, this work indicates that the spatial heterogeneity in grazing intensity has a decisive role to play in the functioning of water-limited ecosystems.
	\end{abstract}
	\maketitle	
	\section{Introduction}
\label{intro}
Dry-lands, consisting of arid and semi-arid region and the dry subtropics, make up roughly 41\% world's land mass and support approximately 38\% human population \cite{reynolds2007global}. Almost 10 to 20\% of these areas face acute land degradation and unfortunately this percentage is expected to grow due to global warming \cite{reynolds2007global,gowda2018signatures}. Dry-land ecosystem has become a key priority for ecologists due to rising concerns about desertification and biodiversity loss, and corresponding effect on ecosystem functioning. Plants and vegetation constitute the underlying energy base of all trophic levels and thus play a crucial role in keeping these water limited ecosystems afloat. 
\par
Varying water stress often results in triggering self-organization of spatial vegetation patterns \cite{von2001diversity}. Since the pioneering work Ref. \cite{macfadyen1950vegetation} in 1950, a substantial number of studies have documented large scale spatial vegetation patterns in dry-lands via aerial photographs and satellite images \cite{deblauwe2012determinants,tlidi2018observation,fernandez2019spiral}. Apart from regular labyrinthine grass patterns in arid or semi-arid landscapes, irregular patterns like groves within grasslands or spots of uncovered ground within a grass matrix, have been observed worldwide \cite{deblauwe2008global,martinez2014minimal,fernandez2014strong,escaff2015localized,fernandez2019front,parra2020formation,clerc2021localised}. These patterns are considered to be the key to understand the processes responsible for desertification, and these understandings can be utilized to combat the catastrophic effect of climate change \cite{dawes2016localised,mau2015reversing}. The emerging characteristics of the self-organization mechanisms govern the system engineering at ecosystem levels like primary or secondary
production, flexibility against increasing environmental pressure, stability \cite{bastiaansen2018multistability,guttal2007self}. Due to the vast spatial and temporal scale of formation of these patterns, mathematical models have become the primary tool in studying vegetation dynamics (see Ref. \cite{zelnik2013regime} for review). 
Several researchers attributed this self-organization of patterns to different processes: positive feedback between vegetation biomass and water infiltration \cite{hillerislambers2001vegetation}, competition among vegetation patches for ground-water due to uptake by roots \cite{von2001diversity,gowda2014transitions}, non-local water uptake by laterally extended roots and enhancement of root system with biomass growth \cite{gilad2004ecosystem,meron2007localized}, or plant-plant interaction only \cite{lefever1997origin,tlidi2020interaction}. To incorporate the dispersal of plants and water movement, several models based on partial differential equations have been proposed \cite{hillerislambers2001vegetation,gilad2007mathematical,messaoudi2020patchy}. One such model that have been a subject of several extension in last two decade is Klausmeier model \cite{klausmeier1999regular} (which will be detailed in next section) due to its lucidity and fundamental nature \cite{eigentler2018analysis,eigentler2019metastability,sun2021mathematical,xue2020spatiotemporal,kabir2022numerical}. This plant-water model is basically a reaction-diffusion-advection system in which the water-uptake feedback loop and spatial displacement of plant-water, have been taken care of in a minimalistic approach. 
\par 
Foraging by herbivores has long been identified as one of the principle influencer on the physiognomy, structure, and functioning of vegetation, ranging from landscape scale to a single plant systems \cite{hobbs1996modification,noy1975stability}. Grazing ecosystems supply large amounts of consumable protein, however too much anthropogenic foraging activities makes ecosystem increasingly vulnerable to stressful environments. Several studies \cite{mueller2014patterns,rietkerk1997alternate} have emphasized that vegetation patterns and foraging by herbivores are interlinked and are sensitive to degradation under extreme environmental events, like droughts. However, most of the prevailing spatial models on dry-land ecosystem have considered the loss of vegetation due to herbivore grazing in a marginalized way; grazing have been thought to be proportional to the vegetation density and modeled by adding a constant term to the plant senescence. But, empirical studies have revealed that foraging by herbivores depends on several factors like spatial distribution of vegetation, quality of forage, behavior of grazer \cite{focardi1996ungulates}. A greater portion of grazers gets attracted by places with higher vegetation concentration, thereby resulting in in-homogeneous grazing pressure. Recently, in an elegant approach, Siero et al. \cite{siero2018nonlocal,siero2019grazing} have derived a grazing term that incorporates the underlying effect of spatially heterogeneous vegetation distribution over foraging and implemented it in the generic plant-water model by Klausmeier \cite{klausmeier1999regular}. \par 
In Ref. \cite{siero2018nonlocal,siero2019grazing},  Siero et al. have used mean-density dependent response term to showcase the dependence of grazing pressure at any position on vegetation elsewhere. This approach intrinsically presumes herbivory at any particular location to depend equally on vegetation everywhere. However, in real scenario the distance between grazer and vegetation also plays a decisive role \cite{siero2018nonlocal,gordon2008ecology,worton1989kernel}. Navigation to the forage, i.e. detecting and traveling to the food item, depends on the characteristic of the grazer, like sight and olfactory cues \cite{gordon2008ecology,roese1991habitat}. Hence, the grazing strategy of a herbivore will be more dependent on availability of vegetation within a certain range, rather than the whole domain. To this end, in this current work, the influence of vegetation over grazing at any particular location is considered to vary with the intermediate distance. This work primarily focuses on addressing two interconnected questions: (1) What impact does this change bring to the self-organization of patterns? (2) Does it have any effect on the system-response towards increasing aridity? Here the mean density is replaced by a spatially weighted average density, using convolution integral with a normalized weight function, that estimates the utilization of space for grazing as a function of the distance. Moreover, a generalized nature is maintained throughout the analysis, so that further modification can be made by using data-driven weight functions.    
\par
In Sect. \ref{Model Description}, the working model is proposed after describing the Klausmeier model in detail and then a thorough mathematical analysis is carried out in Subsection \ref{Stability Analysis}. In Sect. \ref{Response To Changing in Precipitation Level}, numerical simulations are carried out to evaluate the resilience of the system toward varying environmental pressure. In a parsimonious approach, two particular types of weight functions are chosen here for having a comparative understanding and differentiating the obtained results from earlier studies \cite{siero2018nonlocal,siero2019grazing}. Finally, all the findings are ecologically interpreted in Sect. \ref{Discussion} and few potential extensions to this model are suggested.
\section{Model Description and Analyses}\label{Model Description and Analysis}

\subsection{Model description}\label{Model Description}

This section first reviews the original model introduced by Klausmeier, which is then extended by incorporating non-local grazing to the equation for vegetation. Then, Subsection \ref{Model Analysis} presents the systematic mathematical analysis of the modified system. \par
For modeling dry-land vegetation patterns on sloped terrain, Klausmeier \cite{klausmeier1999regular} proposed a reaction-diffusion-advection system:

\begin{align}\label{KL}
\begin{cases}
\begin{split}
&\frac{\partial W}{\partial T} = A-LW-RWN^2+V \frac{\partial W}{\partial X_1},\\ \\
&\frac{\partial N}{\partial T} = RJWN^2-MN+ D\nabla^2 N,
\end{split}
\end{cases}
\end{align}
where $\nabla^2 =  \frac{\partial^2 }{\partial X_{1}^2}+\frac{\partial^2 }{\partial X_{2}^2}$ is the Laplacian operator. $W(T; \vec{X})$ and $N(T; \vec{X})$ are the density of surface water and plant biomass respectively at location $(T;\vec{X}) \in {(T>0) \times \mathbb{R}^2 }$. The first equation of system (\ref{KL}) depicts the dynamics of water; here the source term $A$ is the uniform rate at which water is added via precipitation. The term $-LW$ accounts for the water-loss because of evaporation and the advective term $V\frac{\partial W}{\partial X_1}$ symbolizes the downhill movement of water along the sloping ground. The term $RWN^2$ corresponds to the water-uptake by the plant-roots, here the appearance of non-linearity is a consequence of the positive feedback between plant growth and water seepage. 
The second equation presents vegetation dynamics, where $RJWN^2$ presents the  plant growth; $J$ being the yield of vegetation biomass per unit consumption of water. The term $-MN$ specifies the loss of vegetation due to natural death and grazing of plants by herbivores. Lastly the diffusion $ D\nabla^2 N$ is for modeling the spatial spread of vegetation by means of seed dispersal or clonal growth. \par 
While considering banded vegetation on a slope, the Klausmeier model (\ref{KL}) fails to produce stationary patterns in flat land. To this end several researchers \cite{zelnik2013regime,kealy2012nonlinear,siteur2014beyond} have omitted the advective term (i.e. $V=0$). Further they had added a soil water diffusion term $E\nabla^2 W$ to the first equation of system (\ref{KL}) to encapsulate the movement of surface water due to the spatial heterogeneity in infiltration rate \cite{rietkerk2002self} :
\begin{align}\label{KL_modified}
\begin{cases}
\begin{split}
&\frac{\partial W}{\partial T} = A-LW-RWN^2+E\nabla^2 W,\\ \\
&\frac{\partial N}{\partial T} = RJWN^2-MN+ D\nabla^2 N.
\end{split}
\end{cases}
\end{align}
\par
In model system (\ref{KL}) and (\ref{KL_modified}), senescence of vegetation biomass is assumed to be independent of the spatial distribution of vegetation and grazing by herbivores has been considered as included in this linear term $MN$. But empirical studies \cite{focardi1996ungulates,fryxell2004predictive} on semi arid region tells that grazing depends on spatial distribution of vegetation; herbivores get attracted more to superior forage. This leads to spatially heterogeneous grazing pressure unlike system (\ref{KL_modified}) where it is constant. To incorporate this phenomena, Siero et al. \cite{siero2018nonlocal,siero2019grazing} considered model (\ref{KL_modified}) in one spatial dimension and modified it by adding a density dependent response function $\tilde{G}$. Moreover, the grazing pressure at any spatial point depends on vegetation density not only at that point, but also elsewhere. Taking this non-locality in account they have taken $\tilde{G}$ to be a function of mean vegetation density $\tilde{N}$, where 
\begin{align}\label{mean}
\tilde{N}:=\frac{1}{|\Omega|}\int_{\Omega}N(z)dz,
\end{align} 
$|\Omega|$ denotes the length of the spatial domain $\Omega \in \mathbb{R}$ under consideration. Three different type grazing functions have been used, the first type coincides with existing notion of linear mortality in Klausmeier model, i.e. grazing pressure is considered to be constant: $\tilde{G}=\tilde{M}_{loc}$. Type II is for sustained grazing (e.g. livestock farming), foraging is optional in this case as food shortage will be reimbursed by supplementary food; i.e. demographic response is kept at a constant level all the time. Considering that herbivores maintain a saturating functional response (Holling type II), the grazing function is given by $\tilde{G}=\tilde{M}_{sus}/(\tilde{K}_{sus}+\tilde{N})$, where $\tilde{K}_{sus}$ is the half persistence level. The third type is for natural scenario where grazing is obligatory for survival, only a section of grazers will be able to survive by acquiring sufficient amount of food and rest will disappear. Approximating this demographic response by a sigmoid function (Holling type III), the resulting grazing pressure becomes $\tilde{G}=\tilde{M}_{nat}\tilde{N}/(\tilde{K}_{nat}^2+\tilde{N}^2)$. The resulting model is given by:
\begin{align}\label{KL_graz}
\begin{cases}
\begin{split}
&\frac{\partial W}{\partial T} = A-LW-RWN^2+E\nabla^2 W,\\ \\
&\frac{\partial N}{\partial T} = RJWN^2-(M+ \tilde{G}(\tilde{N}))N+ D\nabla^2 N,
\end{split}
\end{cases}
\end{align} 
where $\nabla^2 =  \frac{\partial^2 }{\partial X^2}$ is the one-dimensional Laplacian operator.
One noteworthy fact is that the non-local term appears non-linearly in the system. For ease of mathematical analysis, system (\ref{KL_graz}) is non-dimensionalised with the following substitutions \cite{siero2018nonlocal,siero2019grazing}:
\begin{align*}
&	w = WJ\sqrt{\frac{R}{L}},\qquad n=N\sqrt{\frac{R}{L}},\quad x=X\sqrt{\frac{L}{D}},\qquad t=TL,\\
&e=\frac{E}{D},\qquad	a=\frac{AJ}{L}\sqrt{\frac{R}{L}},\qquad m_0 = \frac{M}{L},\qquad m_{loc}=\frac{\tilde{M}_{loc}}{L},\\
&  m_{sus/nat}=\frac{\tilde{M}_{sus/nat}}{L}\sqrt{\frac{R}{L}},\qquad K_{sus/nat}=\sqrt{\frac{R}{L}}\tilde{K}_{sus/nat}.
\end{align*}  
The dimensionless model is given by 
\begin{align}\label{final}
\begin{cases}
\begin{split}
&\frac{\partial w}{\partial t} = a-w-wn^2+e\frac{\partial^2 w}{\partial x^2},\\[5pt]
&\frac{\partial n}{\partial t} = wn^2-(m_0+ G(\tilde{n}))n+ \frac{\partial^2 n}{\partial x^2}.
\end{split}
\end{cases}
\end{align}
where 
\begin{align}\label{KL_graz_fun}
G(\tilde{n})=
\begin{cases} 
\begin{aligned}
&m_{loc} &&\text{for local grazing (type I)},\\[5pt]
&\frac{m_{sus}}{K_{sus}+\tilde{n}} &&\text{for sustained grazing (type II)},\\[5pt]
&\frac{m_{nat}\tilde{n}}{K_{nat}^2+\tilde{n}^2} &&\text{for natural grazing (type III)}.
\end{aligned}
\end{cases}
\end{align}
\par  

In this study, system (\ref{final}-\ref{KL_graz_fun}) is taken as base model. One inherent assumption in this model is that, grazing by herbivores at any particular spatial point depends equally on vegetation densities at all other points. But, ecological studies on browsing and grazing have showed that the rate of consumption of vegetation biomass depends on the perceptual abilities (e.g. sight, olfaction) of the animal \cite{gordon2008ecology,roese1991habitat}. When the observational area is vast (which is the case for most of the empirical studies on dry-land vegetation dynamics), the grazing strategy of a herbivore will be more dependent on availability of vegetation nearby, rather than the whole domain. To capture this ecological fact more realistically, the definition of mean vegetation density (\ref{mean}) is modified:
\begin{align}\label{mean_new}
\tilde{n}(x,t)=\int \rho(|x-x^{\prime}|)n(x^{\prime},t)dx^{\prime},
\end{align}
where $\rho(x)$ is a normalized kernel function (i.e. $\int \rho (|y|)dy=1$) accounting for the weighted mean vegetation density. The use of absolute value $|x-x^{\prime}|$ in definition (\ref{mean_new}) ensures spatial isotropy of the kernel function. In a parsimonious representation, here it is assumed that the influence of vegetation density at any position, over the grazing decreases with the distance from the grazer; that is why the weight function $\rho$ is considered to be a monotonically decreasing function of distance in this study. In the next subsection \ref{Model Analysis}, at first the model (\ref{final}-\ref{KL_graz_fun}-\ref{mean_new}) will be analyzed mathematically in a general setting and then a Gaussian distribution function will be used as a particular case for numerical simulation here. Moreover, periodic boundary conditions will be taken to lessen the effect of the boundary and to mimic an infinite domain.

\subsection{Model analysis}\label{Model Analysis}
\subsubsection{Homogeneous steady states}\label{Uniform States}
\begin{figure*}[t]
	\centering
	
	\begin{minipage}{0.32\linewidth}
		\centering
		\includegraphics[width=\linewidth]{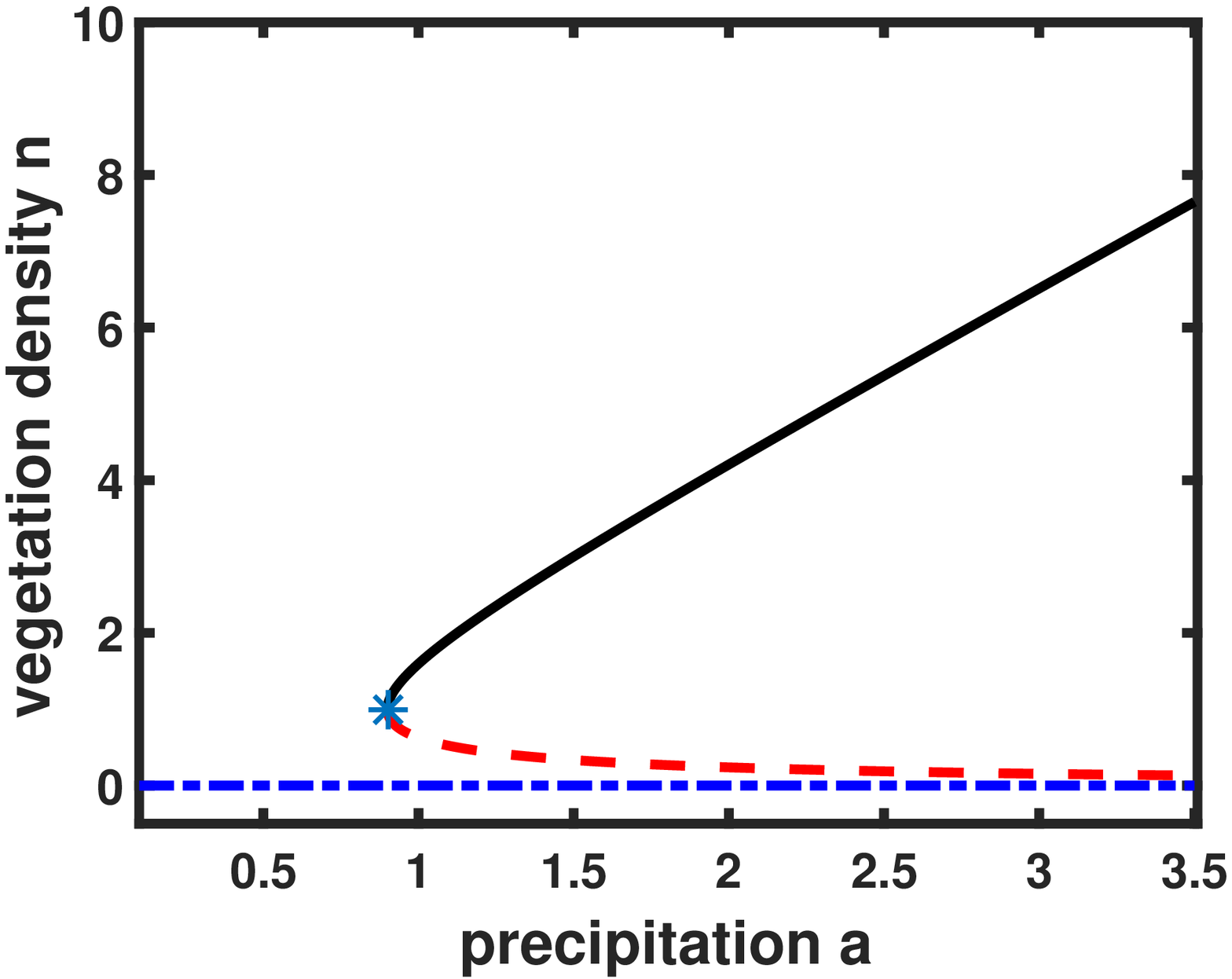}
		\subcaption{} 
		\label{Fig1a}
	\end{minipage}
	\hfill 
	\begin{minipage}{0.32\linewidth}
		\centering
		\includegraphics[width=\linewidth]{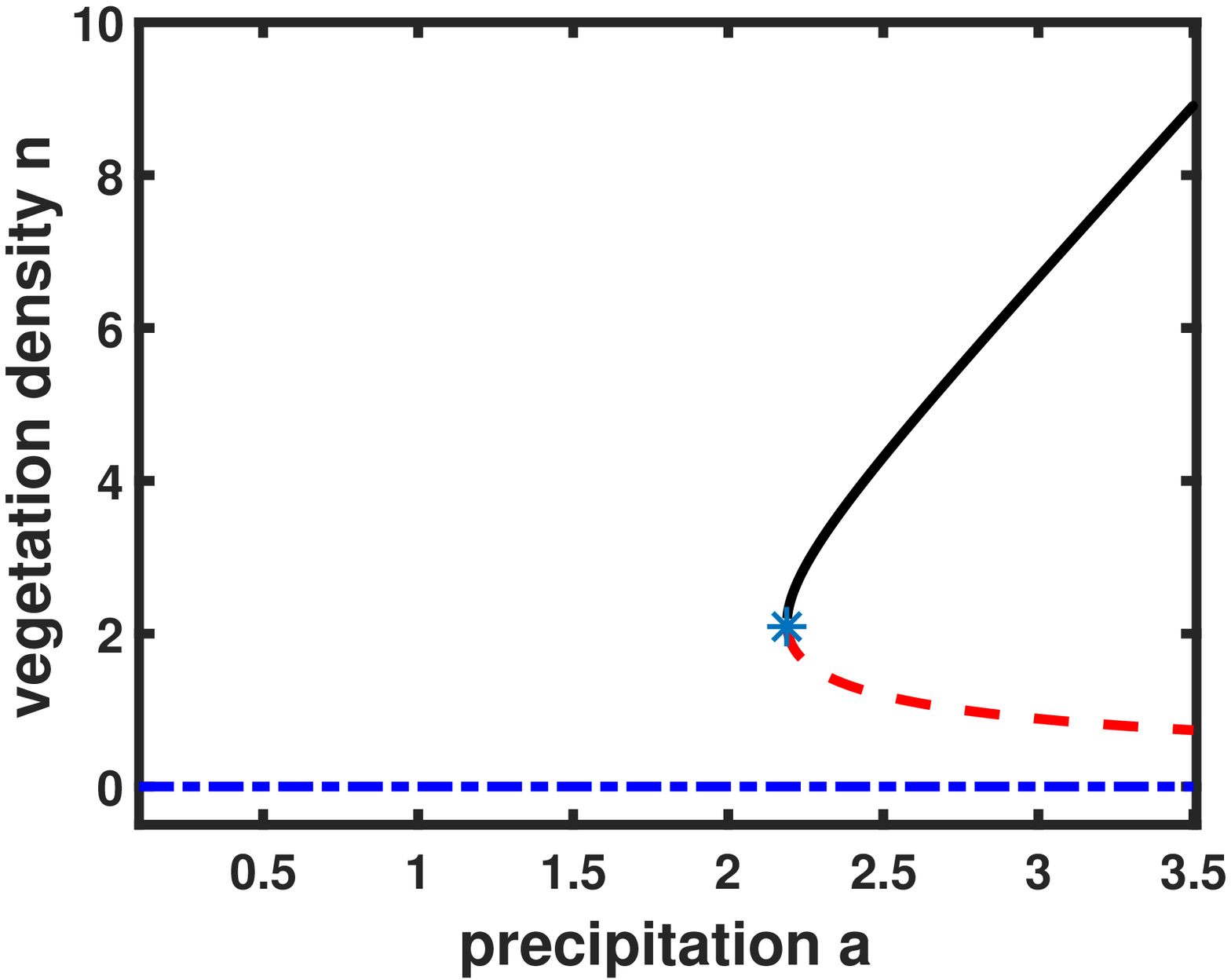} 
		\subcaption{}
		\label{Fig1b}
	\end{minipage}
	\hfill
	\begin{minipage}{0.32\linewidth}
		\centering
		\includegraphics[width=\linewidth]{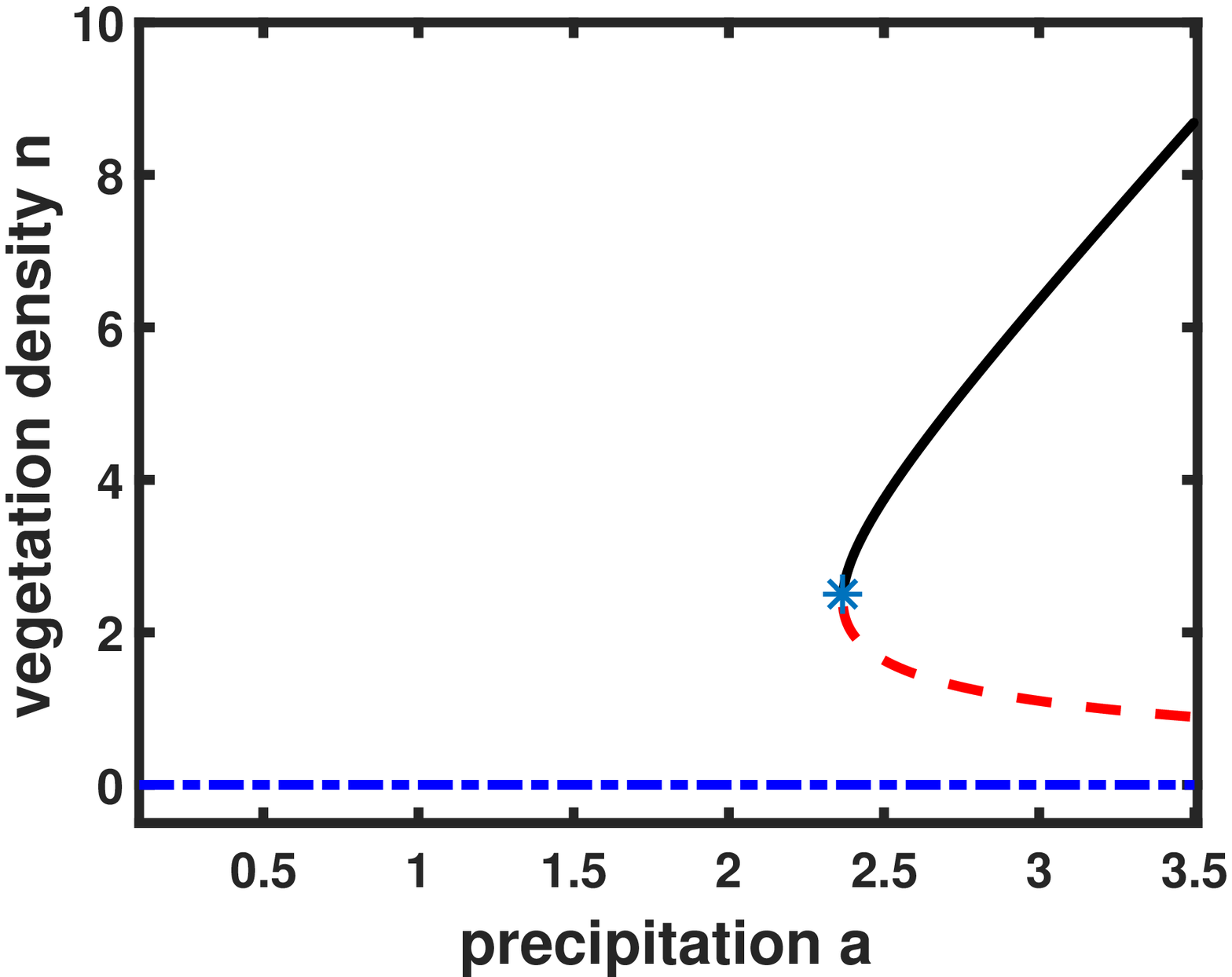}
		\subcaption{} 
		\label{Fig1c}
	\end{minipage}
	\caption{Existence of HSSs and their stability against spatially homogeneous perturbation: chained blue line(\protect\chainblue) is the bare state $\mathcal{B}$ (always stable); solid black line (\protect\blackline) and dashed red line (\protect\dashedred) are the stable and unstable branch of $\mathcal{V}$, respectively. The * denotes the critical transition from one state to another, known as Tipping point. Parameters are same as Ref. \cite{siero2019grazing}: $e=500, m_0=0.225,m_{loc}=0.225,  m_{sus}=m_{nat}=1.5, K_{sus}=K_{nat}=0.3$.   (a) Local Grazing, (b) Sustained grazing, (c) Natural grazing}
	\label{Fig1}
\end{figure*}

This subsection summarizes the existence of homogeneous steady states (hereafter, HSSs) of system (\ref{final}-\ref{KL_graz_fun}). For a homogeneous steady state, $\frac{\partial }{\partial t}w(t,x) = \frac{\partial }{\partial t}n(t,x) =0$ and $\frac{\partial^2  }{\partial x^2}w(t,x)=\frac{\partial^2 }{\partial x^2} n(t,x)=0$. Using these in system (\ref{final}-\ref{KL_graz_fun}),
\begin{subequations}
	\begin{align}
	&a-w-wn^2=0,\label{condA1}\\[5pt]
	&n(wn-m_0-G(\tilde{n}))=0.\label{condA2}
	\end{align}
\end{subequations}
Hence there are two possibilities for HSS: a bare state ($\mathcal{B})$ with $w=a,n=0$ everywhere; and homogeneously vegetated state ($\mathcal{V})$ with $w = \frac{a}{1+n_*^2},n=n_*>0$, where $n_*$ satisfies 
\begin{align}\label{nonzero_eq_eqn}
wn_*-m_0-G(\tilde{n}_*)=0.
\end{align} 
It is noteworthy that there can be more than one  homogeneously vegetated state ($\mathcal{V})$ depending on parameter values and type of grazing function. The main focus of this article is on the sustained and natural grazing type, as the local grazing case have already been studied extensively \cite{sherratt2007nonlinear}. 
\paragraph{Local grazing}
As $G(\tilde{n})=m_{loc}$ is constant in local grazing, it is basically same as the pre-existing models that considers linear mortality.
Apart from the bare state ($\mathcal{B})$, there also exist two homogeneous steady states $(w_S,n_S)$ and $(w_N,n_N)$, when $a>2(m_0+m_{loc})$ \cite{sherratt2010pattern},
\begin{align*}
&w_S=\frac{2(m_0+m_{loc})^2}{a-\sqrt{a^2-4(m_0+m_{loc})^2}},\\[5pt] &n_S=\frac{a-\sqrt{a^2-4(m_0+m_{loc})^2}}{2(m_0+m_{loc})};\\[5pt]
& w_N=\frac{2(m_0+m_{loc})^2}{a+\sqrt{a^2-4(m_0+m_{loc})^2}},\\[5pt] &n_N=\frac{a+\sqrt{a^2-4(m_0+m_{loc})^2}}{2(m_0+m_{loc})}.
\end{align*}   
These two HSSs originate from a saddle-node bifurcation at  $a=2(m_0+m_{loc})$.  Linear stability analysis against homogeneous perturbation \cite{sherratt2005analysis} reveals that the bare state $\mathcal{B}$ is always stable (for $m_0+m_{loc}>0$, which is always true). Moreover, $(w_S,n_S)$ is unstable (actually saddle) and $(w_N,n_N)$ is a node \cite{siteur2014beyond}. 
But for ecologically meaningful parameter values for semi-arid environment, $(w_N,n_N)$ is stable to homogeneous perturbation. The location and stability of steady states against spatially uniform perturbation, has been described in  \autoref{Fig1a}. Furthermore, the linear stability analysis against spatially heterogeneous perturbation \cite{siteur2014beyond} shows that the steady state $(w_N,n_N)$ can lose its stability through Turing bifurcation; although the bare state $\mathcal{B}$ always remains stable.  
\paragraph{Sustained grazing}
In this case, equation (\ref{nonzero_eq_eqn}) takes the form 
\begin{align}\label{nonzero_eq_eqn_sus}
wn-m_0-\frac{m_{sus}}{K_{sus}+\tilde{n}}=0.
\end{align}
Now for uniformly vegetated state $\mathcal{V}$, using the normalization condition of the kernel, we have
\begin{align}\label{n_tilde_sus}
\tilde{n}=\int \rho(|x-x^{\prime}|)n(x^{\prime},t)dx^{\prime}=n_*\int \rho(|x-x^{\prime}|)dx^{\prime}=n_*.
\end{align} 
Substituting $w = \frac{a}{1+n_*^2},n=n_*$ into equation (\ref{nonzero_eq_eqn_sus}) and using (\ref{n_tilde_sus}), a cubic equation for $n_*$ is obtained:
\begin{align}\label{cubic}
\begin{split}
m_0n_*^3 + (m_0K_{sus}-a+m_{sus})n_*^2 +(m_0-aK_{sus})n_* + m_0K_{sus}&\\
+m_{sus}=0&.
\end{split}
\end{align}
Only the positive solutions of (\ref{cubic}) correspond to the physically admissible $\mathcal{V}$ steady states. As all the parameters are non-negative, the number of sign changes in the sequence of this cubic polynomial's coefficients is either zero or two. Moreover, there are no sign change in (\ref{cubic}) when $a<min(m_0K_{sus}+m_{sus},\frac{m_0}{K_{sus}})$, hence Descartes' rule of signs implies that equation (\ref{cubic}) will have no positive solution in that case. Therefore, for sufficiently small level of precipitation, the bare state $\mathcal{B}$ will be the only HSS. Numerical simulation shows that for $a$ greater than some threshold value, there actually exists two branches of homogeneously vegetated steady state $\mathcal{V}$ (See \autoref{Fig1b}).  

\paragraph{Natural grazing}
Proceeding in a similar fashion like sustained grazing, a quartic polynomial for $n_*$ is obtained:
\begin{align}\label{quartic}
\begin{split}
m_{0}n_*^4+(m_{nat}-a)n_*^3+(K_{nat}^2+1)m_{0}n_*^2+ (m_{nat}-aK_{nat}^2)n &\\
+ m_{0}K_{nat}^2=0.&
\end{split}
\end{align}
Depending on model parameters, this quartic equation may or may not have positive solution. If the model parameters are such that $a<min(m_{nat},\frac{m_{nat}}{K_{nat}^2})$,  then the sequence of coefficients of the polynomial (\ref{quartic}) will have no sign change; i.e. there will be no uniformly vegetated state $\mathcal{V}$ for low precipitation level. Numerical methods are used for finding roots of this quartic equation which show that, for the parameter choice Ref. \cite{siero2019grazing} (which was based on earlier studies and empirical data), equation (\ref{quartic}) has two positive zeros when $a$ values are greater than some critical threshold. Consequently for sufficiently large values of $a$, there will be two uniformly vegetated states $\mathcal{V}$ apart from the bare state $\mathcal{B}$ (See \autoref{Fig1c}).   

\subsubsection{Stability analysis }\label{Stability Analysis}
Linear stability analysis is a broadly utilized tool to acquire the temporal evolution of small perturbations to the homogeneous steady states of the system. When given perturbation gets amplified over time, the steady state will be unstable; but if the perturbation decays with time, taking the system back to the homogeneous stationary state, then that steady state will be stable. For notational convenience, let $(w_{eq},n_{eq})$ be a homogeneous stationary state (it can be bare state $\mathcal{B}$ or vegetated state $\mathcal{V}$). If a small perturbation is given to the HSS $(w_{eq},n_{eq})$, the resulting density will be
\begin{align}\label{perturb}
\begin{bmatrix} w(x,t) \\ n(x,t) \end{bmatrix} = \begin{bmatrix} w_{eq} \\ n_{eq} \end{bmatrix}+ \begin{bmatrix}
\epsilon_1\\ \epsilon_2
\end{bmatrix}\psi(x,t)
\end{align}
where $|\epsilon_1|\ll 1$ and $|\epsilon_2|\ll 1$. Now plugging equation (\ref{perturb}) into equation (\ref{mean_new}) and using the normalization condition of the kernel function,
\begin{align*}
\tilde{n}(x,t)&=\int \rho(|x-x^{\prime}|)(n_{eq}+\epsilon_2 \psi(x^{\prime},t))dx^{\prime} \nonumber\\ 
&=n_{eq}+ \epsilon_2 \int \rho(|x-x^{\prime}|)\psi(x^{\prime},t)dx^{\prime}.\\ 
\text{So,}\quad G(\tilde{n})&=G\bigl(n_{eq}+ \epsilon_2 \int \rho(|x-x^{\prime}|)\psi(x^{\prime},t)dx^{\prime}\bigr).
\end{align*}
As the perturbation considered here is very small, the function $G$ can be expanded using Taylor series:
\begin{align}\label{Taylor}
G(\tilde{n})= G(n_{eq}) + \epsilon_2 G^\prime (n_{eq}) \int \rho(|x-x^{\prime}|)\psi(x^{\prime},t)dx^{\prime}+ O(\epsilon_2 ^2)
\end{align}
where $ G^\prime(n_{eq})= \bigl(\frac{d G}{d \tilde{n}}\bigr)_{\tilde{n}=n_{eq}}$.\\
Now substituting (\ref{perturb}) into the first equation of system (\ref{final}) and neglecting the higher order terms in $\epsilon_1$ and $\epsilon_2$ we have,
\begin{align}\label{linearised_fisrt_raw}
\begin{split}
\epsilon_1 \frac{\partial \psi}{\partial t}=\epsilon_1 e \frac{\partial^2 \psi}{\partial x^2}-\epsilon_1(1+n_{eq}^2)\psi-\epsilon_2 2w_{eq}n_{eq}\psi &\\ +(a-w_{eq}-w_{eq}n_{eq}^2)&.
\end{split}
\end{align}
As $(w_{eq},n_{eq})$ is homogeneous steady state of system (\ref{final}), we have $a-w_{eq}-w_{eq}n_{eq}^2=0$. Hence (\ref{linearised_fisrt_raw}) becomes
\begin{align}\label{linearised_first}
\epsilon_1 \frac{\partial \psi}{\partial t}=\epsilon_1 e \frac{\partial^2 \psi}{\partial x^2}-\epsilon_1(1+n_{eq}^2)\psi-\epsilon_2 2w_{eq}n_{eq}\psi.
\end{align}
Substituting (\ref{perturb}) and (\ref{Taylor}) into the second equation of system (\ref{final}) and neglecting higher order terms in $\epsilon_1$ and $\epsilon_2$ we have,
\begin{align}\label{linearised_second}
\epsilon_2 \frac{\partial \psi}{\partial t}=& \epsilon_1 n_{eq}^2 \psi +\epsilon_2 \biggl\{\frac{\partial^2 \psi}{\partial x^2}+ \bigl(-m_0+2w_{eq}n_{eq}-G(n_{eq})\bigr)\psi \nonumber\\
&	 - n_{eq} G^\prime (n_{eq}) \int \rho(|x-x^{\prime}|)\psi(x^{\prime},t)dx^{\prime}\biggr\}
\end{align}
\par 
Applying Fourier transform to this linearized integro-differential system (\ref{linearised_first}-\ref{linearised_second}) for the evolution of $\psi$, we have

\begin{align}\label{pert_matrixform_pre}
\begin{bmatrix} \epsilon_1 \\ \epsilon_2 \end{bmatrix}\frac{\partial \hat{\psi}(k,t)}{\partial t} = L \begin{bmatrix} \epsilon_1 \\ \epsilon_2 \end{bmatrix} \hat{\psi}(k,t),
\end{align}
where
\begin{align}\label{linearised_matrix}
L=
\begin{bmatrix}
-ek^2-1-n_{eq}^2 & -2w_{eq}n_{eq}\\[4ex]
n_{eq}^2 & -k^2 -m_0+2w_{eq}n_{eq}-G(n_{eq})\\& -n_{eq} G^\prime (n_{eq})\hat{\rho}(k)
\end{bmatrix}.
\end{align}
$k$ is known as wave-number, $\hat{\psi}(k,t)= \int exp(ikx)\psi(x,t)dx$ is the Fourier transform of the perturbation and similarly $\hat{\rho}(k)$ is the Fourier transform of the kernel.\par
Now, considering $\hat{\psi}(k,t) \propto exp(\lambda(k)t)$, equation (\ref{pert_matrixform_pre}) becomes

\begin{align}\label{pert_matrixform}
\bigl(L-\lambda(k) I \bigr)  \begin{bmatrix} \epsilon_1 \\ \epsilon_2 \end{bmatrix} = \begin{bmatrix} 0 \\ 0 \end{bmatrix}.
\end{align}
Non-zero solution of the linear system (\ref{pert_matrixform}) exists if and only if \begin{align*}
Det(L-\lambda(k) I) =0,
\end{align*}
which yields a quadratic equation in $k^2$ for the linear growth rate $\lambda$:
\begin{align}
\lambda^2 - Trace(L)\lambda + Det(L)=0 \label{dispersion}.
\end{align}
Equation (\ref{dispersion}) is known as the dispersion relation. Linear stability of the HSSs depends on the essential spectrum (which consists of the two solutions $\lambda_1,\lambda_2$ of the dispersion relation) of the linearized system. $(w_{eq},n_{eq})$ is linearly stable if $\operatorname{Re}(\lambda_1)\leq0$ and  $\operatorname{Re}(\lambda_2)\leq0$, and unstable if either these two inequalities gets altered.
\par
Here stability against both spatially uniform perturbation and heterogeneous perturbation will be considered.
\paragraph{Spatially homogeneous perturbation} 
Note that, the case of spatially homogeneous perturbation corresponds to the wave-number $k=0$. Then the linearized matrix (\ref{linearised_matrix}) takes the form,
\begin{align}\label{linearised_matrix_uniform}
\begin{bmatrix}
-1-n_{eq}^2 & -2w_{eq}n_{eq}\\[4ex]
n_{eq}^2 &  -m_0+2w_{eq}n_{eq}-G(n_{eq})- n_{eq} G^\prime (n_{eq})
\end{bmatrix}.
\end{align}
If $(w_{eq},n_{eq})$ is taken to be the bare state $\mathcal{B}(a,0)$, then (\ref{linearised_matrix_uniform}) becomes, 
\begin{align}\label{linearised_matrix_uniform_bare}
L_{\mathcal{B}}=
\begin{bmatrix}
-1 & 0\\[4ex]
0 &  -m_0-G(0)
\end{bmatrix}.
\end{align}
From (\ref{linearised_matrix_uniform_bare}), it can be concluded that dispersion relation (\ref{dispersion}) for $\mathcal{B}$ will always have negative roots, i.e. bare state $\mathcal{B}$ is linearly stable against spatially homogeneous perturbations for all three grazing types. \par 
Now, if we consider $(w_{eq},n_{eq})$ to be the uniform vegetated state $\mathcal{V}$, then it will satisfy condition (\ref{nonzero_eq_eqn}). So equation (\ref{linearised_matrix_uniform}) takes the form:
\begin{align}\label{linearised_matrix_uniform_vegetated}
L_{\mathcal{V}}=
\begin{bmatrix}
-1-n_{eq}^2 & -2w_{eq}n_{eq}\\[4ex]
n_{eq}^2 &  w_{eq}n_{eq}- n_{eq} G^\prime (n_{eq})
\end{bmatrix}.
\end{align}
Unlike the case for bare state, the sign of the real parts of $\lambda_1,\lambda_2$ (which are actually the eigenvalues of matrix $L_\mathcal{V}$) depend on the value of $w_{eq},n_{eq}$ and the grazing function $G$. Using numerical schemes, the existence and linear stability of these HSSs have been evaluated for all the three types of grazing (\autoref{Fig1}). Observe that up to a certain threshold value of precipitation $a$, bare state $\mathcal{B}$ is the only HSS available. But at some critical value of $a$, saddle-node bifurcation occurs and a stable and an unstable (saddle) branch of uniformly vegetated state $\mathcal{V}$ appear. 

\paragraph{Spatially heterogeneous perturbation}
Putting $w_{eq}=a$ and $n_{eq}=0$ in matrix (\ref{linearised_matrix}), it can be observed that the linearized matrix for bare state $\mathcal{B}$ always have negative eigenvalues, i.e. bare state $\mathcal{B}$ is stable against heterogeneous perturbations also. \par
So far, all the mathematical analysis have been carried out without specifying the explicit form of kernel function. But, from the linearized matrix (\ref{linearised_matrix}), it is evident that the linear stability of the state $\mathcal{V}$ against heterogeneous perturbation, depends on the Fourier transform of the kernel. So, it directly depends on the structure of the kernel function. Without loss of generality, it is assumed that, vegetation at any point have lesser influence over the grazer the further they are; and after certain cut-off distance, the influence is too insignificant to consider. Keeping this assumption in mind, a cut-off Gaussian function \cite{fuentes2004analytical} is used here:  
\begin{align}\label{cut-off Gaussian}
\rho_{1}(x) = \frac{1}{\sigma \sqrt{\pi} \erf(w/\sigma) } exp\biggl[-\biggl(\frac{x^2}{\sigma^2}\biggr)\biggr]\bigl\{\theta[w-x]\theta[w+x]\bigr\}.
\end{align}
Here, $\theta$ is the Heaviside function and $w$ denotes the effective radius of action. The shape of the kernel function is characterized by its width ($\sigma$) and the cut-off length ($w$). For further analysis, the spatial domain is considered to be finite: $[-L,L]$. The cut-off distance $w$ is always within the bound of domain length (i.e. $w\leq L$). The term $\frac{1}{\sigma \sqrt{\pi} \erf(w/\sigma) }$ is the normalization factor, which ensures that the integral of $\rho_{1}(x)$ over the whole domain equals to unity and it serves the purpose of being an averaging function. For this choice of $\rho$, the Fourier transform will be:
\begin{align}
\hat{\rho}_{1}(k) = \frac{exp[-(\sigma k/2)^2]}{2 \erf(w/\sigma)}\biggl[\erf\biggl(\frac{w}{\sigma}+\frac{i k \sigma}{2}\biggr)+\erf\biggl(\frac{w}{\sigma}-\frac{i k \sigma}{2}\biggr)\biggr].
\end{align}
To have a comparative understanding of the role of the kernel function, a second choice of $\rho$ is  considered here:
\begin{align}\label{constant_kernel}
\rho_{2}(x) = \frac{1}{2w}\bigl\{\theta[w-x]\theta[w+x]\bigr\},
\end{align}
and
\begin{align}
\hat{\rho}_{2}(k) = \frac{\sin(kw)}{kw}.
\end{align}
Without loss of generality, both the kernels will be considered with $w=L$ for numerical simulations. In this scenario, $\sigma$ is the parameter which regulates the variation in dependence of grazing intensity over vegetation elsewhere. One noteworthy fact is, for $w=L$, kernel choice $\rho_{2}$ actually corresponds to the uniform weight function, which is the case in Ref. \cite{siero2018nonlocal,siero2019grazing}. This enables us to compare results of this study with the findings of Ref. \cite{siero2018nonlocal,siero2019grazing}. 
\par 
Unlike the case for homogeneous perturbation (i.e. $k=0$), here the growth term $\lambda$ will be dependent not only on the model parameters, but also on the wave number $k$ of 	the perturbation. As the saddle branch of steady state (dashed red part in \autoref{Fig1}) is already unstable against the homogeneous perturbation, only the stable node (solid black part) will be considered here. These stable nodes will get destabilized if the real part of one or both of the solutions $\lambda(k)$ of dispersion relation (\ref{dispersion}) become positive. With the variation in the bifurcation parameter, instability can happen in mainly two ways: Hopf instability (when $Trace(L)$ goes to positive from being negative); and Turing instability (when $Det(L)$ becomes negative from being positive). Numerical simulation reveals that, for realistic choice of ecological parameters \cite{klausmeier1999regular,siero2019grazing}, these stable nodes loose stability only by Turing bifurcation, no Hopf instability is observed in this model. When a uniformly vegetated state is driven out of Turing stability region, spatio-temporal vegetation pattern appears.  
\begin{figure*}
	\centering
	
	\begin{minipage}{0.48\linewidth}
		\centering
		\includegraphics[width=\linewidth]{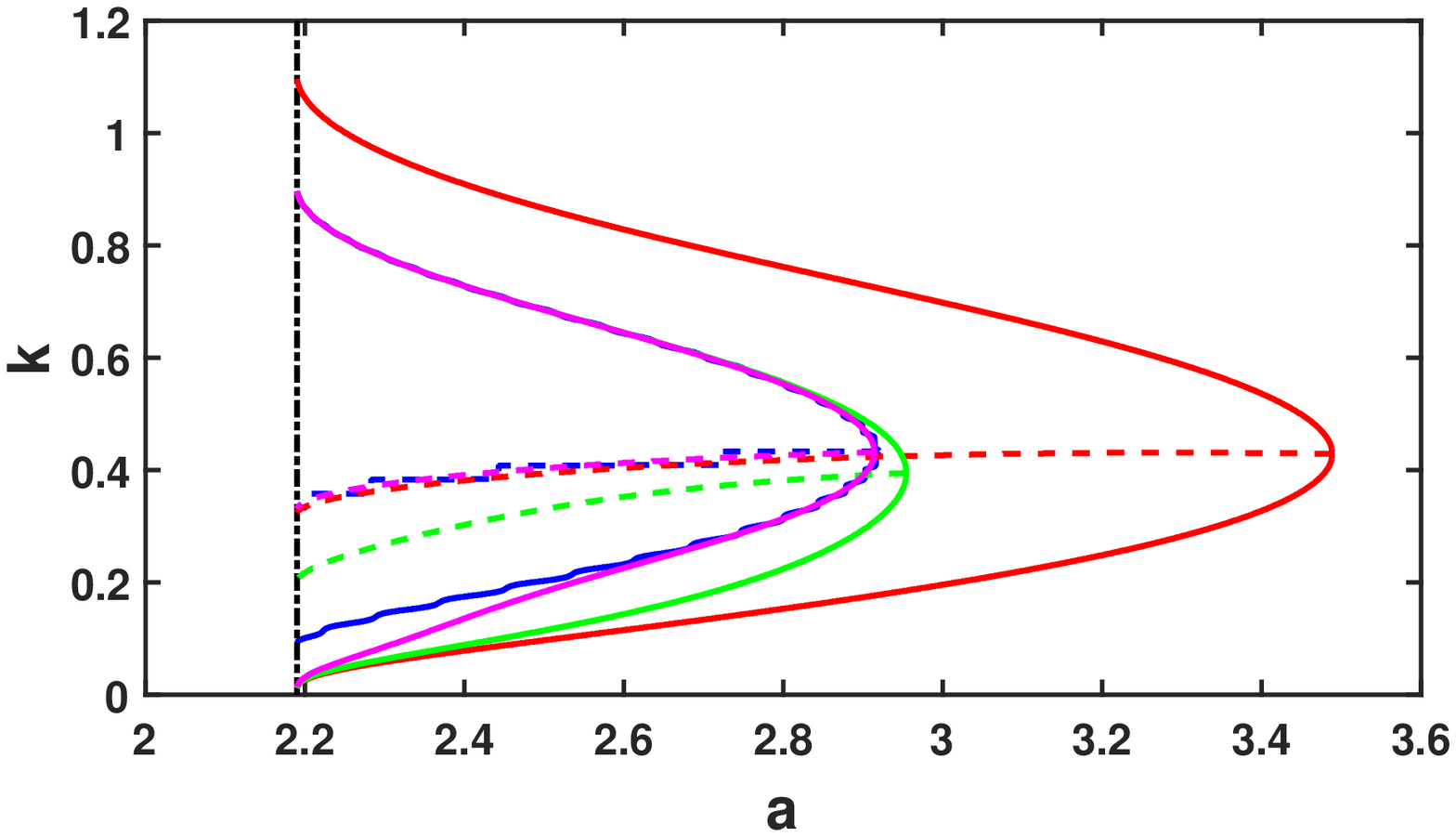}
		\subcaption{} 
		\label{Fig2a}
	\end{minipage}
	\begin{minipage}{0.48\linewidth}
		\centering
		\includegraphics[width=\linewidth]{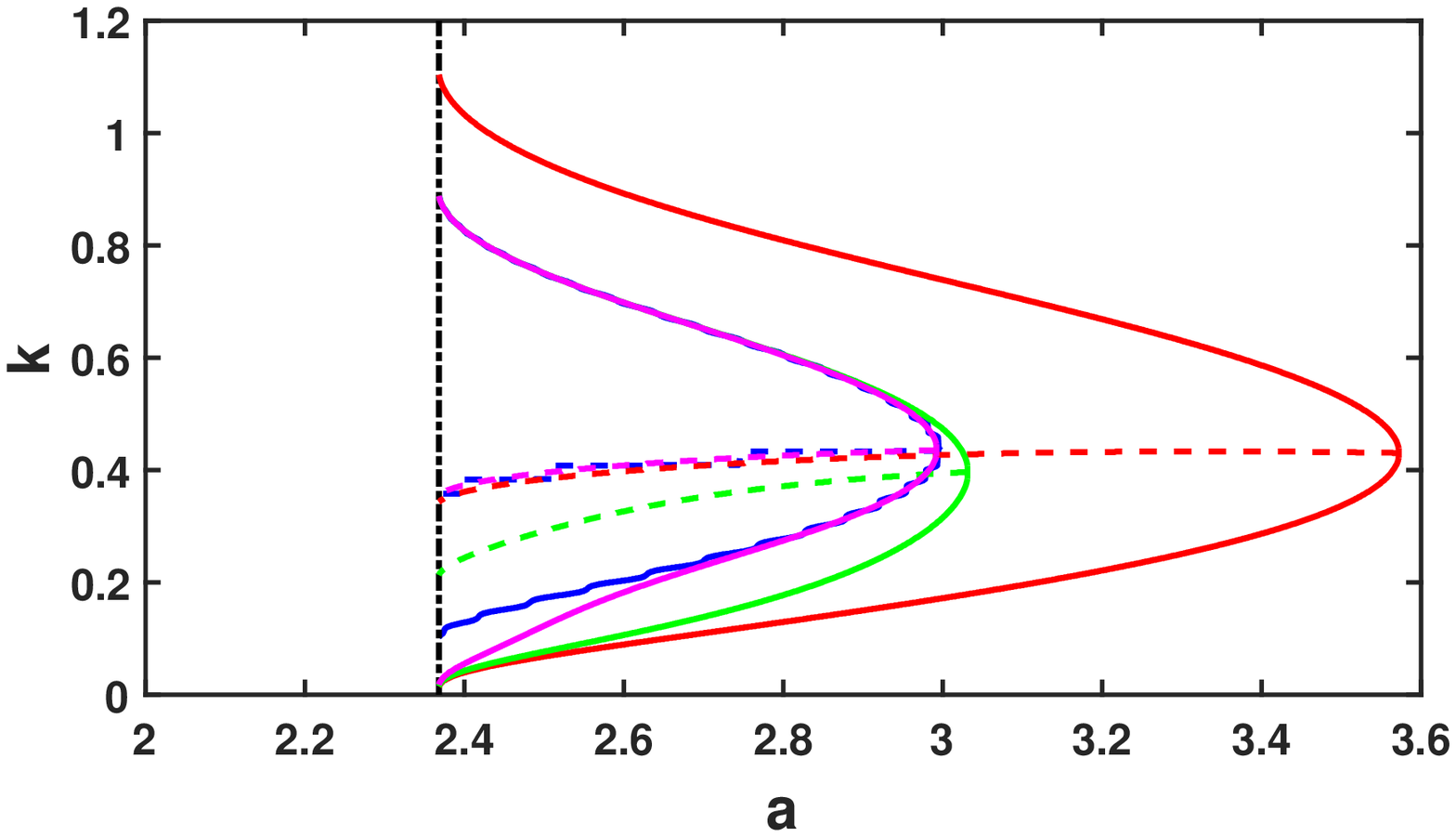} 
		\subcaption{}
		\label{Fig2b}
	\end{minipage}
	
	\caption{Stability scenario against heterogeneous perturbation in $(a,k)$ space. The colored solid lines represent the boundaries of Turing prediction region for different choice of kernel: cut-off Gaussian $\rho_{1}$  having $\sigma=1$ (\protect\redline), $\sigma=8$ (\protect\greenline), $\sigma=15$ (\protect\magentaline); and uniform kernel $\rho_{2}$  (\protect\blueline). The chained line (\protect\chainblack) denotes the lower bound of precipitation, below which the steady state $\mathcal{V}$ ceases to exist 
		. The dashed lines (colored in aforesaid order) represents the mode of perturbations with largest growth rate. Parameters are same as \autoref{Fig1}. (a) Sustained grazing, (b) Natural grazing}
	\label{Turing_prediction_region}
\end{figure*} 

In \autoref{Turing_prediction_region}, values of $a$ at which the homogeneously vegetated state ($\mathcal{V}$) becomes Turing unstable, have been derived. To have a comparative understanding, different structures of kernel function are considered. The black chained line (\protect\chainblack) marks the lower bound of precipitation for which uniform vegetated state ($\mathcal{V}$) persists. Inside the region in $a-k$ space, bounded by the colored solid curves and to the right of the black chained line (\protect\chainblack), the amplitude of given perturbation grows with time, i.e. for every $(a,k)$ pair from this region, the dispersion equation (\ref{dispersion}) will have at least one solution $\lambda$ having positive real part.  This region is termed as Turing prediction region, \cite{siteur2014beyond,siero2015striped} as self-organized spatio-temporal patterns can be expected in this region due to Turing type instability. Moreover, the most unstable mode of the perturbation (i.e. the maximum of $Re(\lambda(a,k))$) for each of $a$ in Turing instability band is derived, and it is denoted by dashed lines (colored in their respective order). When precipitation is getting reduced over time, there will be no pattern formation initially; but at the maximum precipitation value in the Turing prediction region (i.e. the intersection points of dashed and solid curves of same color) the HSS looses stability via Turing bifurcation and spatio-temporal patterns appear. These intersection points will be denoted by $a_T$ hereafter. In other words, the stable branch (black solid curve) in \autoref{Fig1} will end at $a_T$, long before reaching the tipping point, which is basically the level of precipitation at which the non-spatial version of the system shifts from vegetated state $\mathcal{V}$ to the less desirable stable state $\mathcal{B}$. It can also be seen from \autoref{Turing_prediction_region} that the pattern formation starts at less arid situation in case of natural grazing, compared to the sustained grazing. Model runs have confirmed that patterns generated at $a_T$ have a wave-number very close to the $k$ value of these intersection points. Usually in-homogeneous patterns of vegetation, generated when random perturbations are applied to the uniformly vegetated state, are expected to have the mode close to the wave-number of the perturbation that have largest growth rate. 
However, which of these wave-numbers gets chosen at a particular precipitation level, is largely unpredictable and in \autoref{Response To Changing in Precipitation Level} it will be seen that the history of environmental variations has significant role to play in this selection of wave-number.      
\section{Response To Changing Precipitation Level}\label{Response To Changing in Precipitation Level}
It is well-known that the stability characteristics of a system, acquired by linear stability analysis, are generally local in nature, i.e. these predictions holds to be true in a close vicinity to the steady state. Therefore in this section, to have a complete understanding of the dynamical scenario, system (\ref{final}-\ref{KL_graz_fun}) will be studied with gradually varying precipitation parameter. Earlier studies \cite{siteur2014beyond,sherratt2013history} on dry land vegetation revealed that rising environmental pressure (e.g. degrading precipitation level) drives the system to a coarsening cascade of transitions to patterns with increasing wavelengths and for adequately low precipitation level the system goes to bare state. Here, a spatial domain $\Omega = [-250,250]$ (resembling 1 km \cite{siero2018nonlocal}) with periodic boundary conditions will be considered for numerical simulation. All other parameter values (mentioned in \autoref{Fig1}) are in accordance with previous studies \cite{klausmeier1999regular}. Initially for every simulation, the precipitation level is kept at $a=3 (\approx800$  mm  year$^{-1})$, sufficient to support a uniformly vegetated state. Then the precipitation level is set to  decrease at a constant rate $\frac{da}{dt} =  -10^{-4}(\approx 0.1$ mm year$^{-1})$ until $a=0$ is reached \cite{siero2019grazing}. Moreover, to reduce numerical artifacts and incorporate intrinsic noise sources that deterministic equations fail to capture, spatially and temporally uncorrelated multiplicative uniformly
distributed noise of small amplitude $0.05\%$ is added to both the components at every $\frac{1}{4}$ year \cite{siteur2014beyond}. Simulations are carried out in MATLAB using the Backward Time Centered Space (BTCS) difference scheme. To avoid discrepancy that may arise near the boundaries of $\Omega$ during numerical evaluation of the mean density (\ref{mean_new}), both the kernel functions $\rho_{1}$, $\rho_{2}$ are modified accordingly \cite{pal2021effects,pal2020effects}. Following earlier studies \cite{siero2019grazing,siero2018nonlocal}, in the numerical simulations vegetation density $n$ less than $10^{-6}$ has been considered as zero. 

\begin{figure*}
	\centering
	\begin{minipage}{0.48\linewidth}
		\centering
		\includegraphics[width=\linewidth]{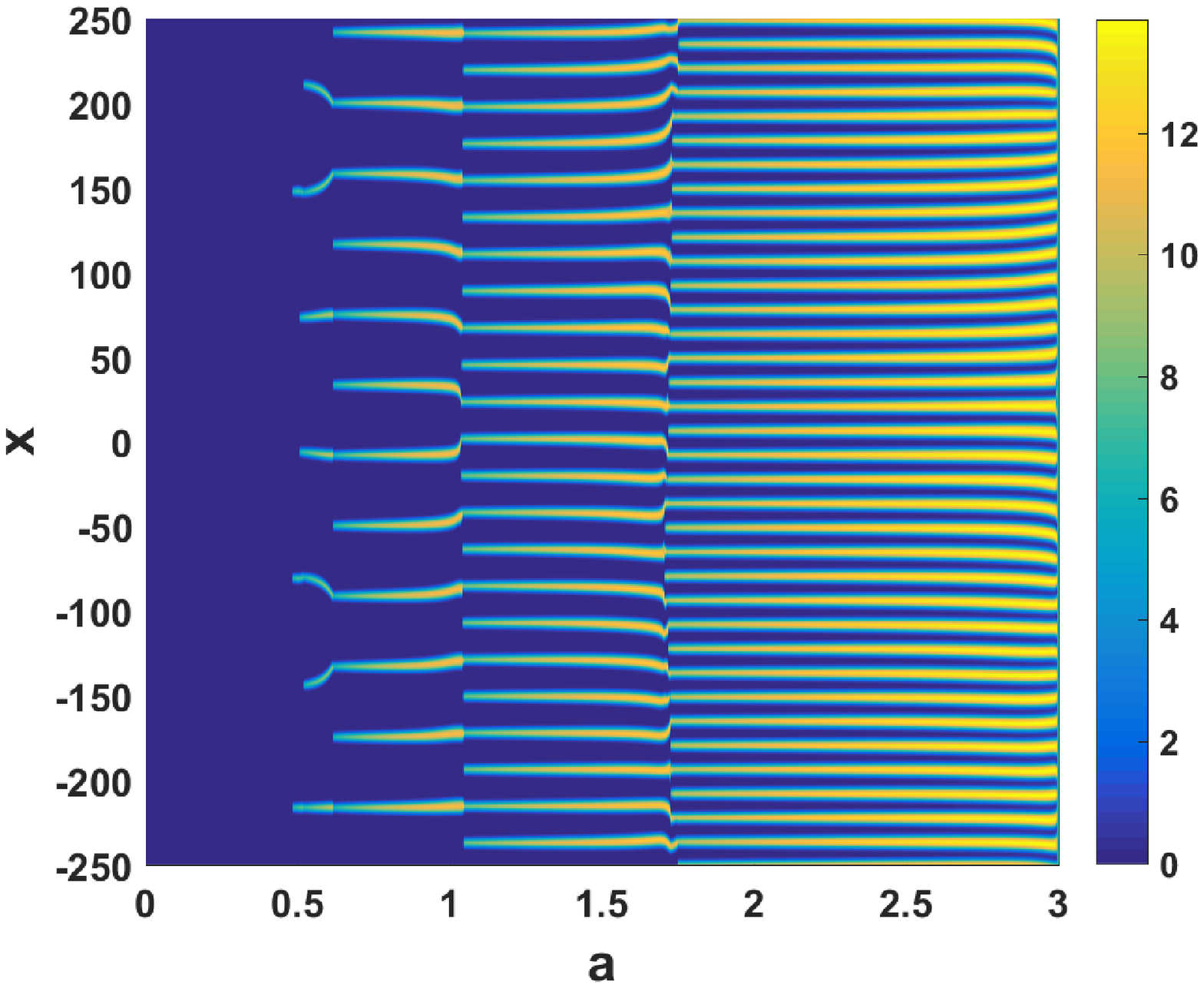}
		\subcaption{} 
		\label{sigma_1_II}
	\end{minipage}
	\begin{minipage}{0.48\linewidth}
		\centering
		\includegraphics[width=\linewidth]{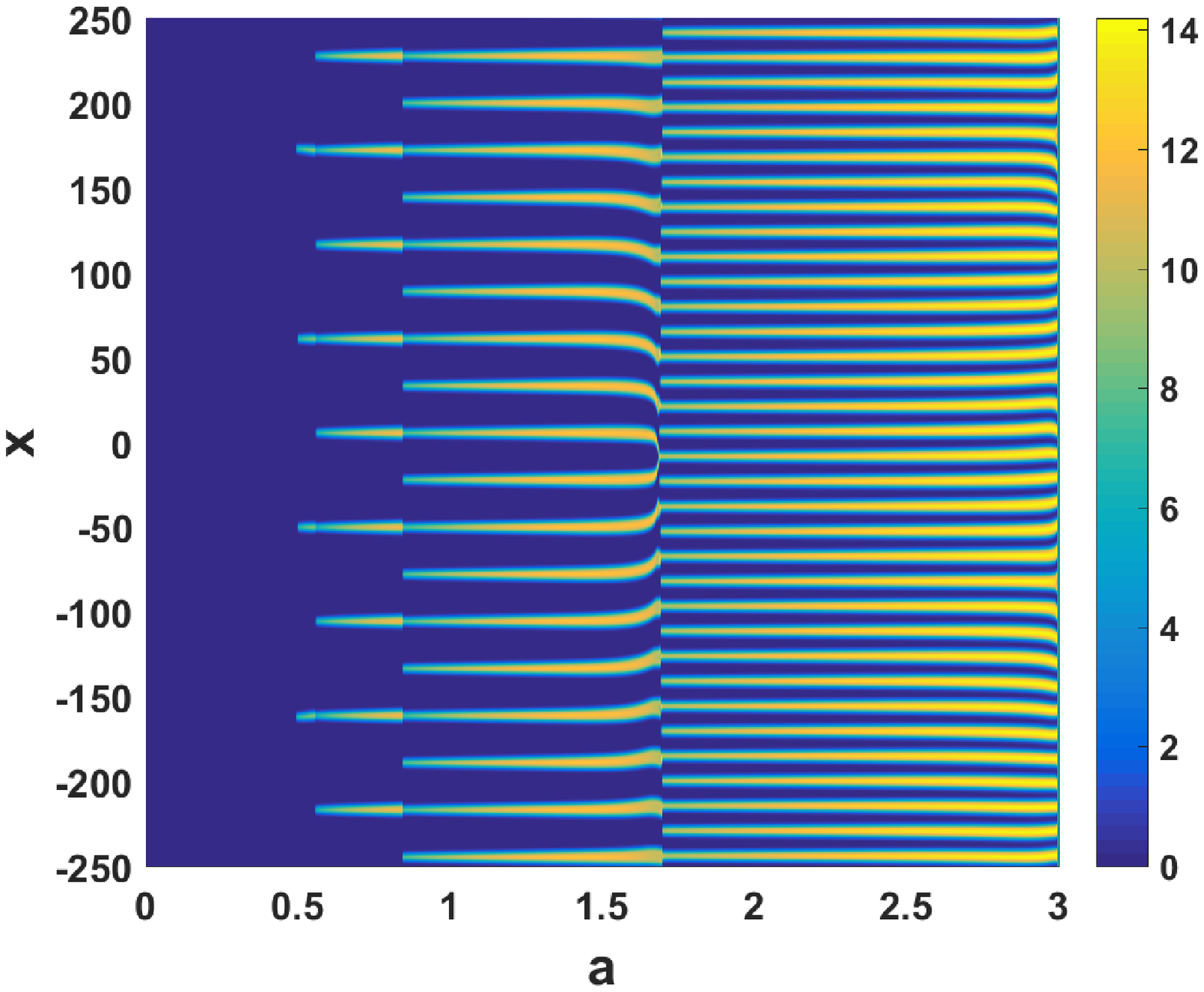} 
		\subcaption{}
		\label{sigma_1_III}
	\end{minipage}
	\newline
	\begin{minipage}{0.48\linewidth}
		\centering
		\includegraphics[width=\linewidth]{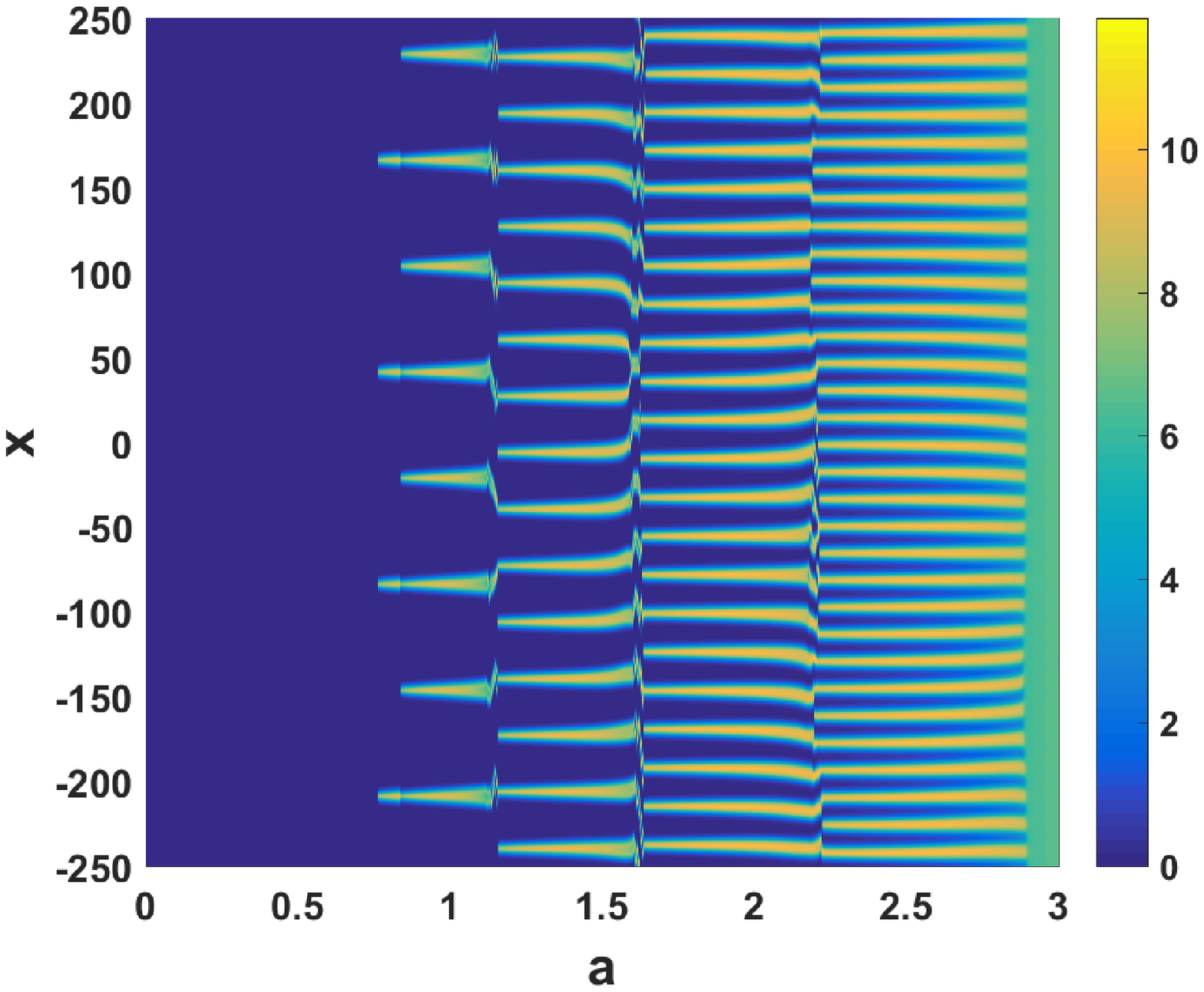}
		\subcaption{} 
		\label{sigma_8_II}
	\end{minipage}
	\begin{minipage}{0.48\linewidth}
		\centering
		\includegraphics[width=\linewidth]{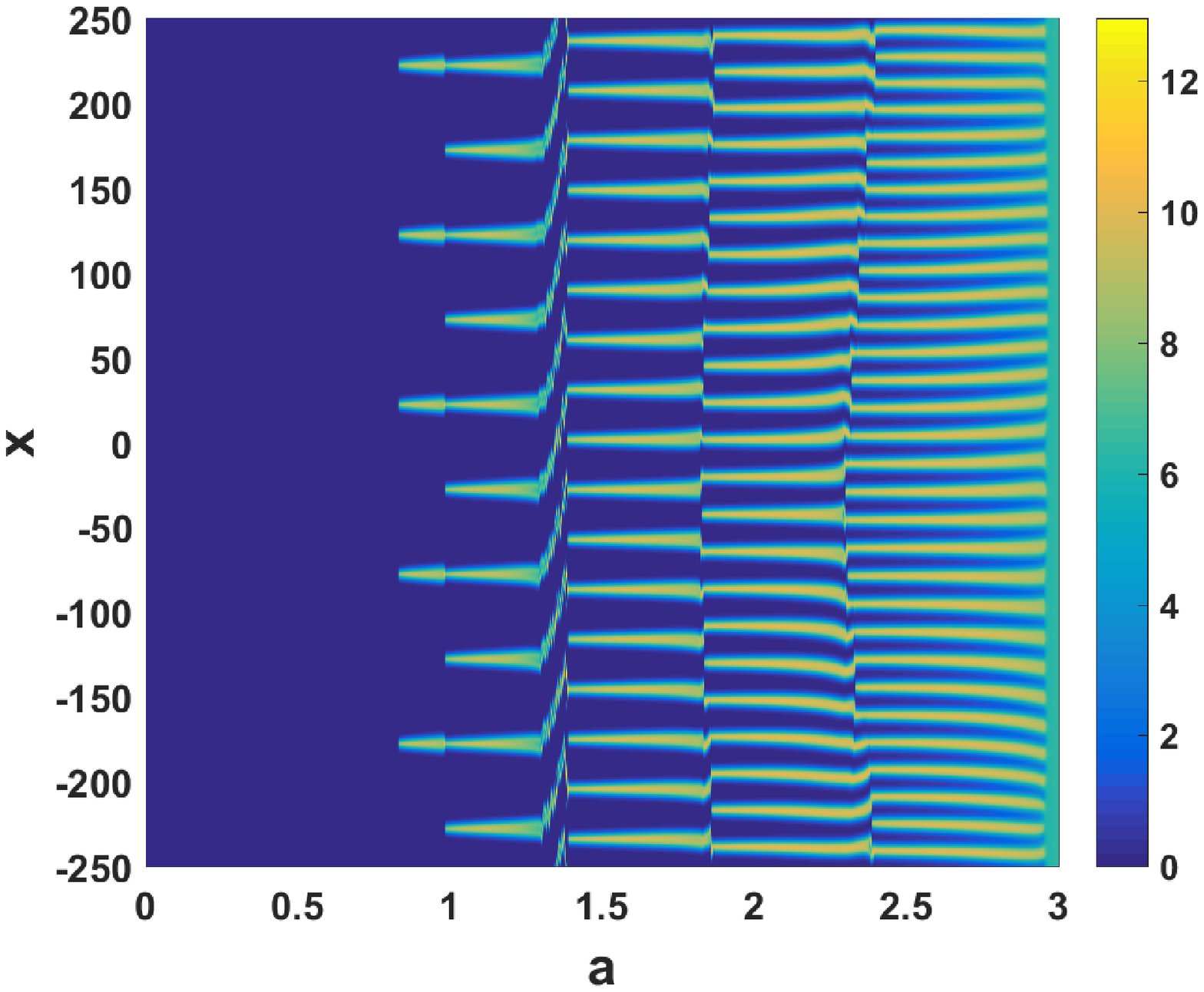} 
		\subcaption{}
		\label{sigma_8_III}
	\end{minipage}
	\newline
	\begin{minipage}{0.48\linewidth}
		\centering
		\includegraphics[width=\linewidth]{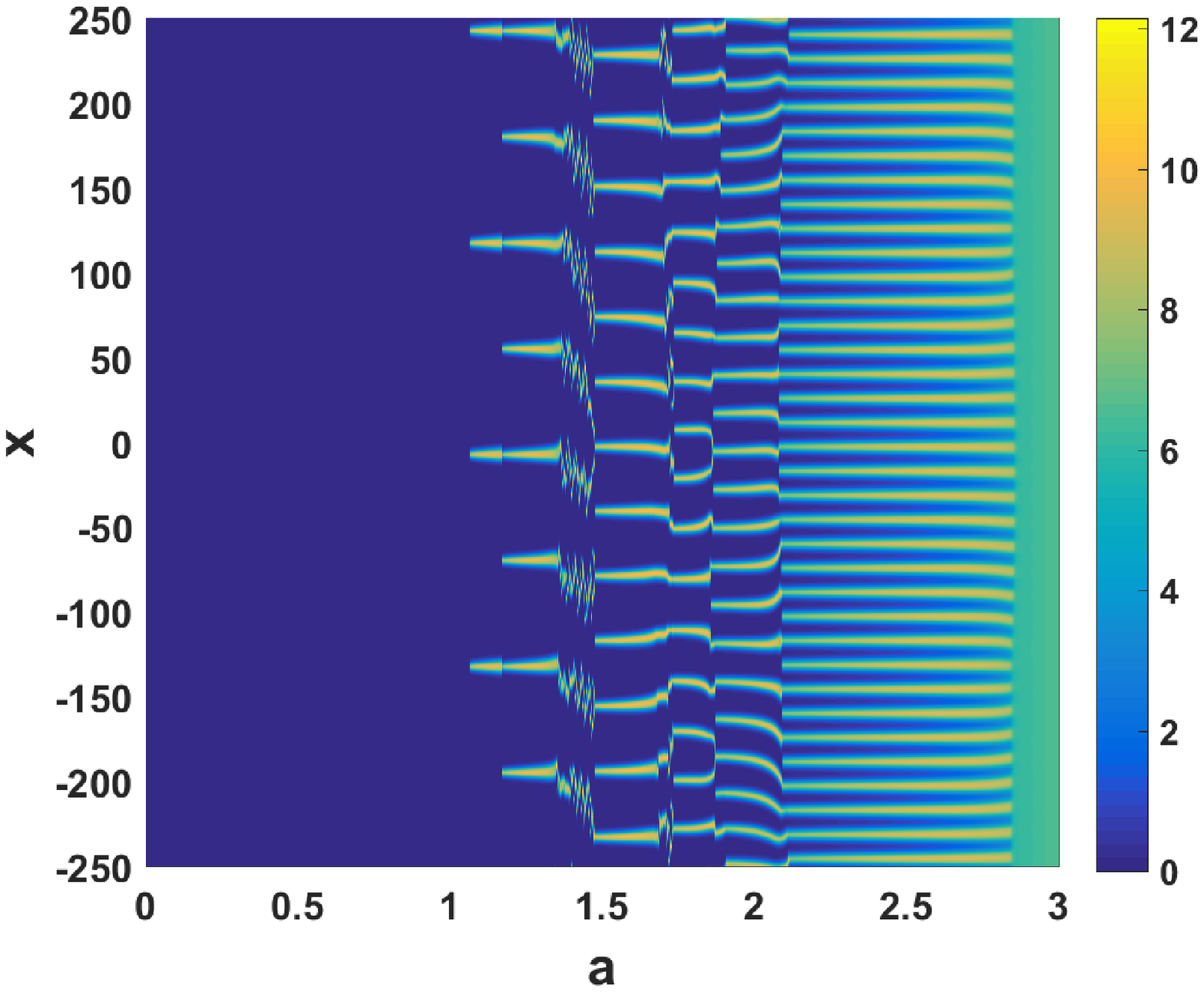}
		\subcaption{} 
		\label{sigma_15_II}
	\end{minipage}
	\begin{minipage}{0.48\linewidth}
		\centering
		\includegraphics[width=\linewidth]{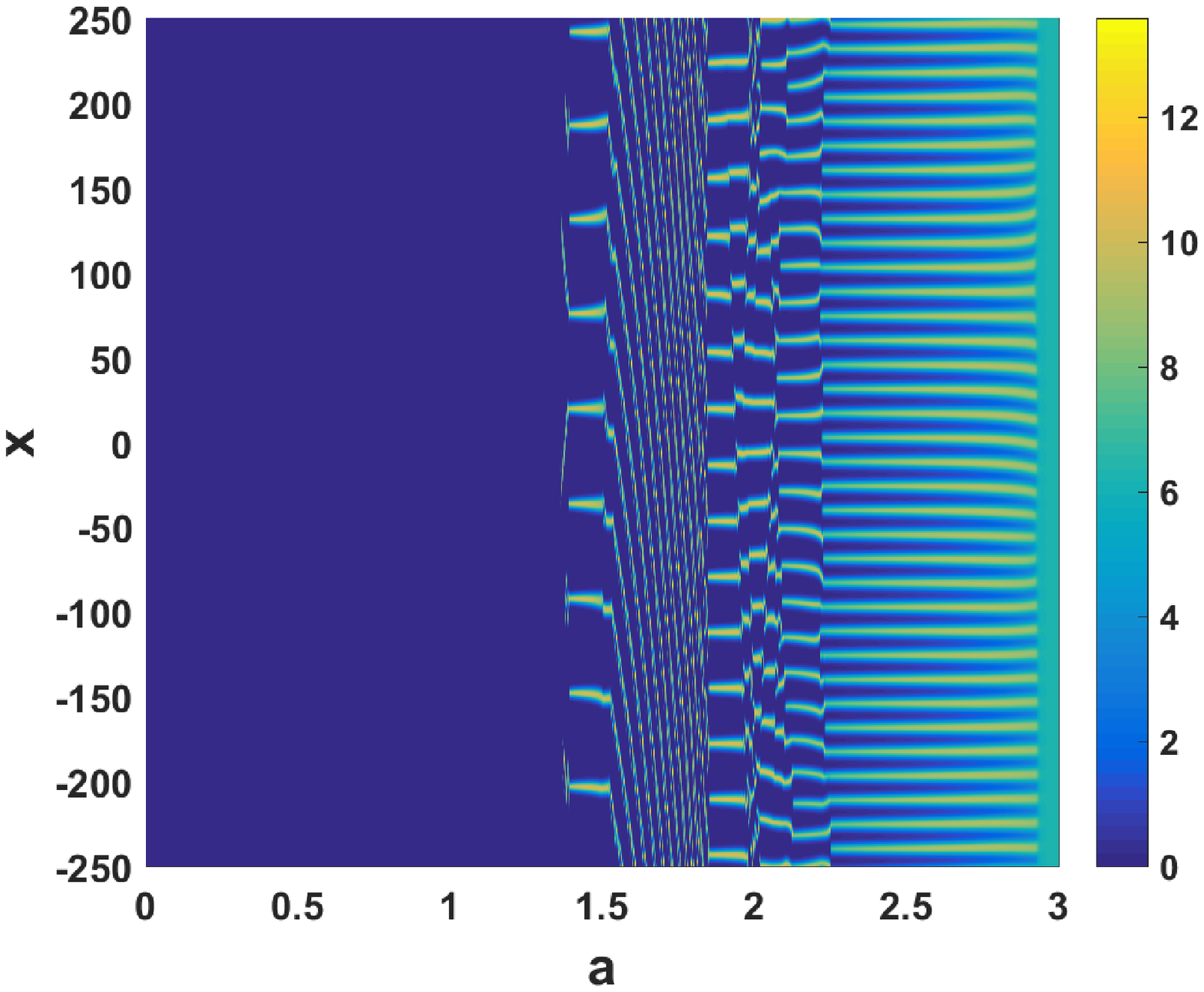} 
		\subcaption{}
		\label{sigma_15_III}
	\end{minipage}
	
	\caption{Simulation results for system (\ref{final}-\ref{KL_graz_fun}) with kernel $\rho_{1}$ (\ref{cut-off Gaussian}) having different $\sigma$ values: (a-b) $\sigma=1$, (c-d) $\sigma=8$, (e-f) $\sigma=15$. The left side column is for Sustained Grazing (a-c-e) and the second column (b-d-f) is for Natural Grazing. In all of these model runs, starting from a homogeneously vegetated state at $a=3$, the precipitation level is set to decrease at a rate $\frac{da}{dt}=-10^{-4}$}
	
	\label{DecreasingRainfall_gaussian}
\end{figure*}

\begin{figure*}
	\centering
	
	\begin{minipage}{0.48\linewidth}
		\centering
		\includegraphics[width=\linewidth]{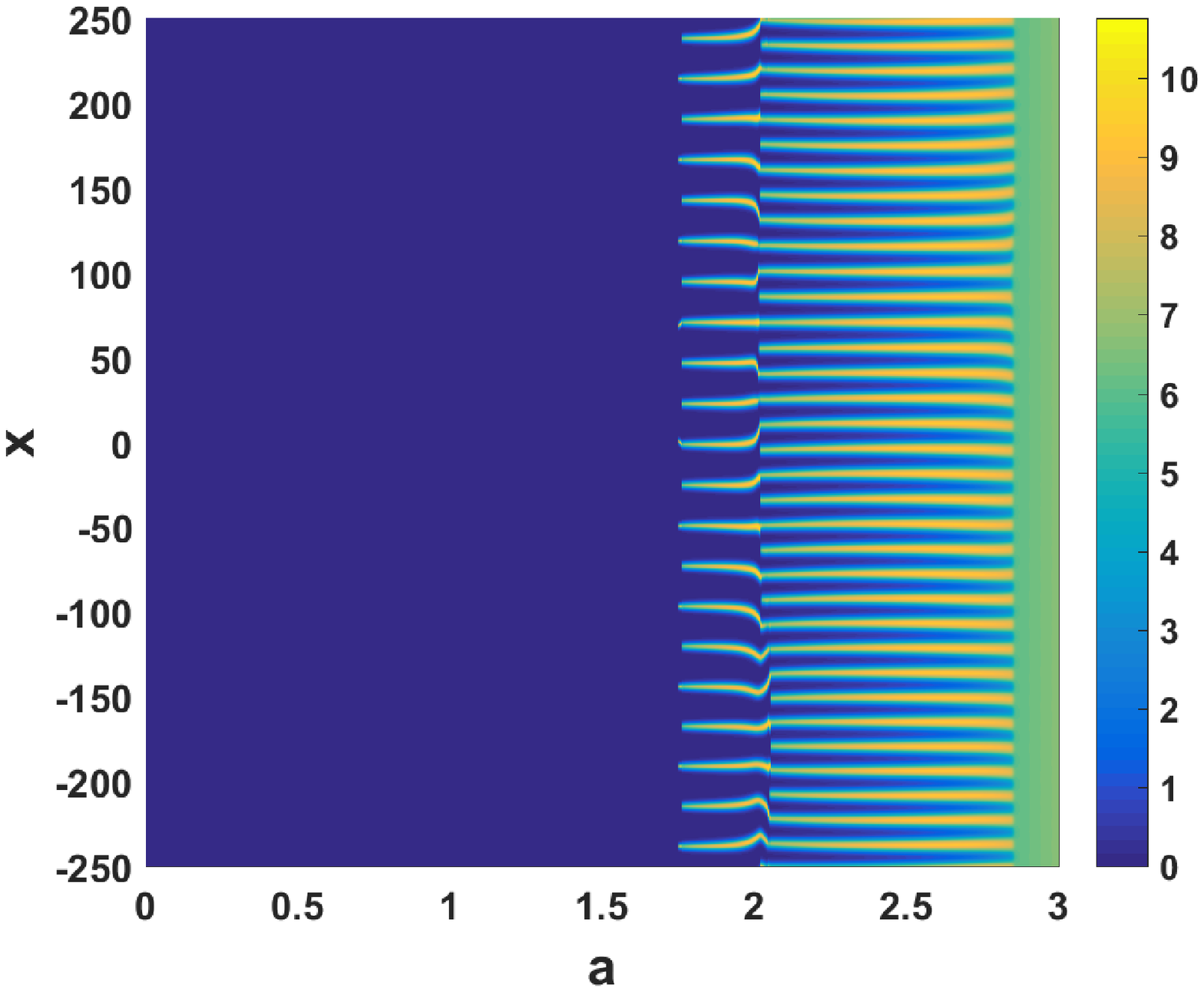}
		\subcaption{} 
		\label{AmNat_2}
	\end{minipage}
	\begin{minipage}{0.48\linewidth}
		\centering
		\includegraphics[width=\linewidth]{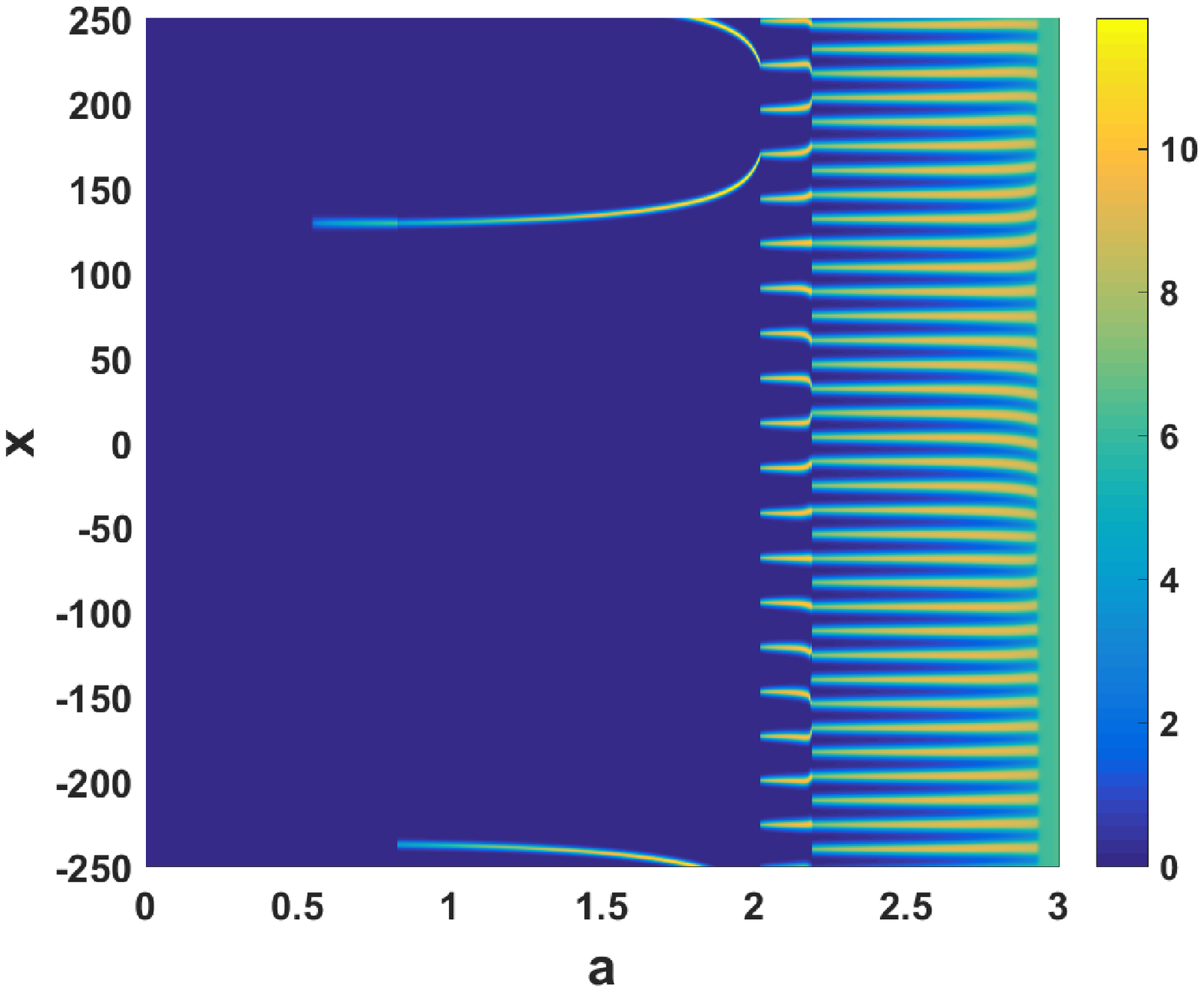} 
		\subcaption{}
		\label{AmNat_3}
	\end{minipage}
	
	\caption{Simulation results for system (\ref{final}-\ref{KL_graz_fun}) with kernel $\rho_{2}$ (\ref{constant_kernel}). The left side column is for Sustained Grazing (a) and the second column (b) is for Natural Grazing. Starting from a homogeneously vegetated state at $a=3$, the precipitation level is set to decrease at a rate $\frac{da}{dt}=-10^{-4}$}
	\label{Decreasing_AmNat}
\end{figure*} 

To showcase the role of kernel function in the system-response (to decreasing precipitation level), model runs have been performed with both cut-off Gaussian $\rho_{1}$ and uniform kernel $\rho_{2}$. Different shapes of kernel $\rho_{1}$ (i.e. $\rho_{1}$ having different widths $\sigma$) have been used in these simulations, results of three of which are shown in \autoref{DecreasingRainfall_gaussian}. It can be seen from these figures that for both type of grazing, as the precipitation reduces, patterns of vegetation appears just after the initial uniformly vegetated state loses its stability through Turing bifurcation. This transition of stability happens at the critical precipitation level $a_T$, mentioned in the previous section. Moreover, wavelengths of these initial patterns matches very closely with the estimation made by linear stability analysis. For example, in case of  $\sigma=8$ with sustained grazing, the bifurcation point in \autoref{Fig2a} is $(a_T,k_T)=(2.954,0.395)$, so the critical wavelength would be $2\pi/k_T \approx15.91$, hence the resulting pattern is expected to have approximately 31 ridges in the computational domain [-250,250], which is exactly the case in \autoref{sigma_8_II}. Furthermore, \autoref{Turing_prediction_region} reveals that the $a_T$ value for $\sigma=1$ is much greater than $a=3$ (which is the starting point of the simulation), so in this case pattern emerges right from the beginning (\autoref{sigma_1_II}-\autoref{sigma_1_III}). But as the width of the weight function $\rho_{1}$ increases, this critical precipitation level $a_T$ decreases. \autoref{sigma_8_II}-\autoref{sigma_15_III} reflect the same phenomena, a homogeneously vegetated state persists until precipitation level reaches $a_T$. However, the rate of lowering of $a_T$ falls off with increment in $\sigma$ and after a certain value (which is $\sigma \approx15$ for aforesaid parameter choice) no more significant change in the value of $a_T$ occurs.  With further decline in precipitation the wavelengths of these patterns remains constant initially, and then undergo number of sudden transitions to patterns with larger and larger wavelength. And after a sufficiently low value of precipitation all of these vegetation patches go extinct and a bare desert state is attained. This desertification threshold varies with the width ($\sigma$) of kernel $\rho_{1}$, as $\sigma$ rises the threshold value of  precipitation for desertification also rises, for both type of grazing. \par 
In \autoref{Decreasing_AmNat} simulations have been performed for $\rho_{2}$ with the same setting as \autoref{DecreasingRainfall_gaussian}. Like the case for $\rho_{1}$, as the precipitation reduces, the initial homogeneously vegetated state undergoes a Turing bifurcation resulting in formation vegetation patterns, and then after a number of pattern transitions a regime shift to desert state occurs. In case of sustained grazing this regime shift is to a state with no vegetation, but a little amount of vegetation continues to exist (the thin strips from $a\approx2$ to $a\approx0.54$ in \autoref{AmNat_3}) for natural grazing, before going to a fully degraded state finally. As expected, these observations matches exactly with the findings of Ref.  \cite{siero2018nonlocal,siero2019grazing}.
\par 
A qualitative comparison between these kernel choices shows that, unlike the kernel $\rho_{2}$, in case of Gaussian kernel $\rho_{1}$, patterned vegetated state goes on till a precipitation level that is very low, in case of sustained grazing also. Moreover, shift toward a completely degraded state happens for much higher value of precipitation in case of $\rho_{2}$ compared to $\rho_{1}$. One more interesting fact is, as the width of kernel $\rho_{1}$ grows, the characteristics of the system become more and more similar to the system with uniform kernel $\rho_{2}$. \par 
\begin{figure*}
	\centering
	
	\begin{minipage}{0.48\linewidth}
		\centering
		\includegraphics[width=\linewidth]{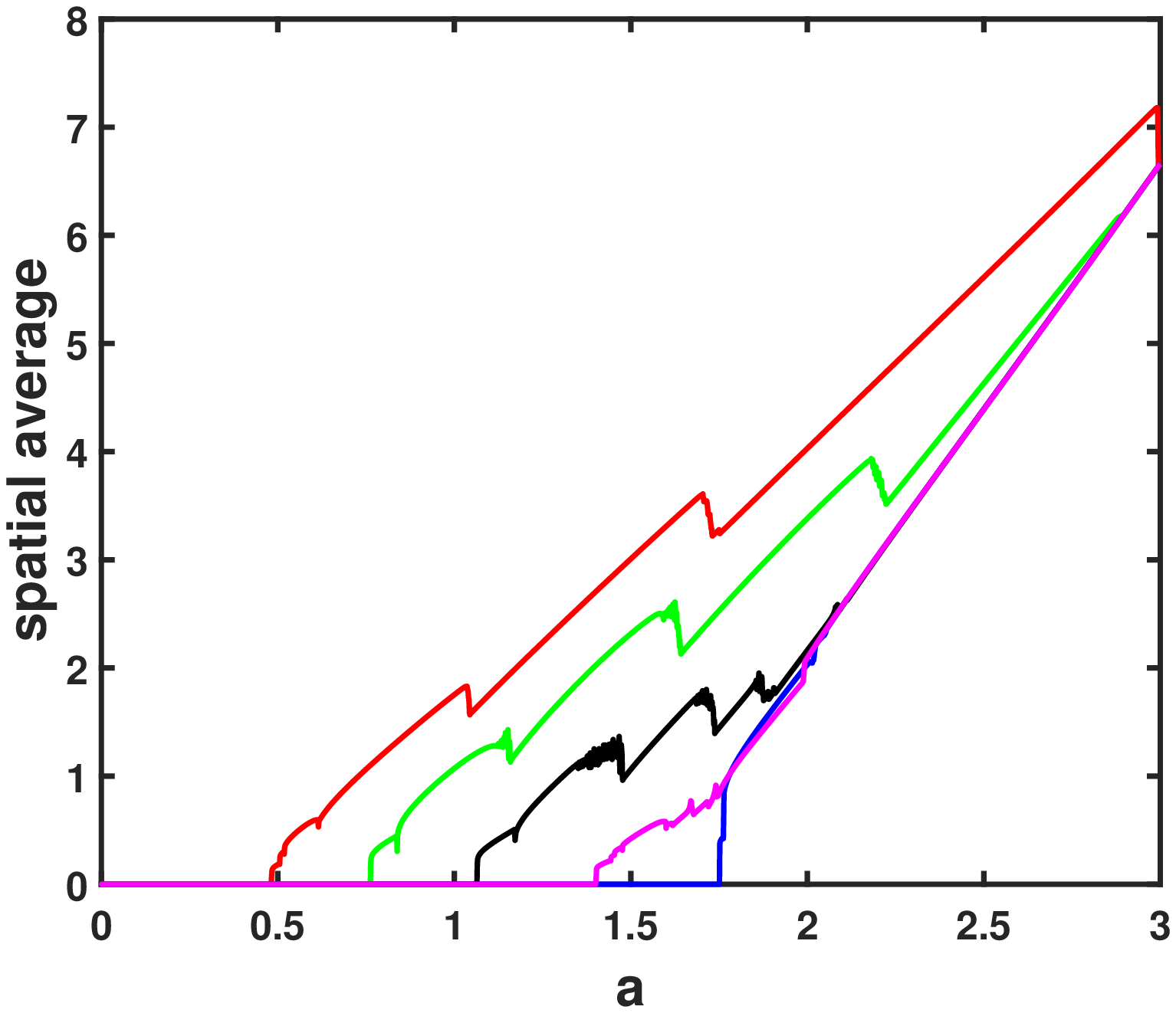}
		\subcaption{} 
		\label{Spatial_Avg_II}
	\end{minipage}
	\begin{minipage}{0.48\linewidth}
		\centering
		\includegraphics[width=\linewidth]{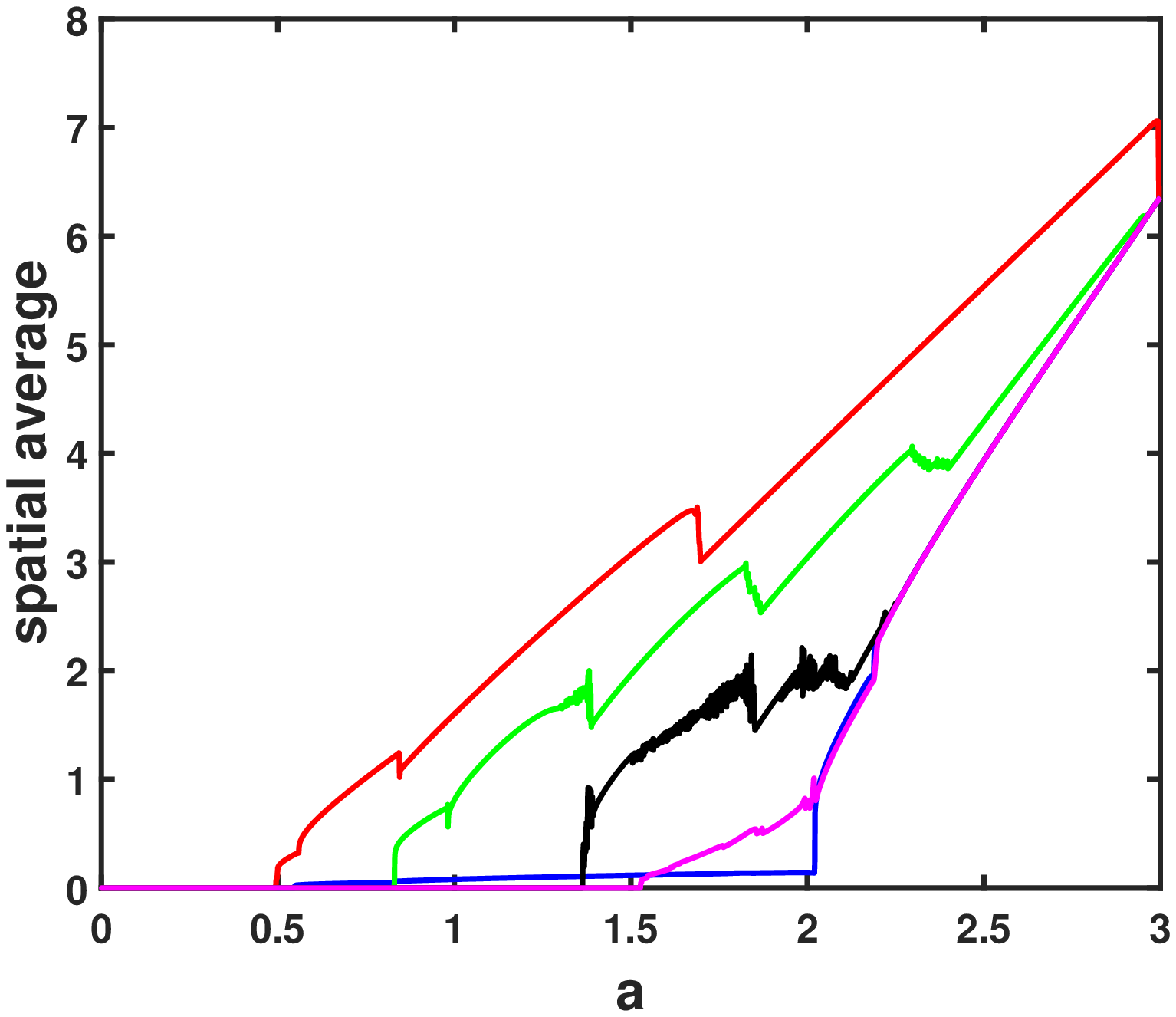} 
		\subcaption{}
		\label{Spatial_Avg_III}
	\end{minipage}
	
	\caption{Spatially averaged vegetation density vs precipitation corresponding to the previous simulations, with cut-off Gaussian $\rho_{1}$  having $\sigma=1$ (\protect\redline), $\sigma=8$ (\protect\greenline), $\sigma=15$ (\protect\blackline), $\sigma=25$ (\protect\magentaline); and uniform kernel $\rho_{2}$  (\protect\blueline). (a) Sustained Grazing  and (b) Natural Grazing}
	\label{Spatial_Avg}
\end{figure*}   
In \autoref{Spatial_Avg}, the spatial averages (i.e. biomass per unit area) of the vegetation biomass are plotted for the aforementioned simulation runs. For both kernel choices, declining precipitation level results in reduction of vegetation biomass. This reduction is not gradual, rather we can see sudden changes in the density curve (\autoref{Spatial_Avg}) whenever the patterns undergoes wavelength adaptations in \autoref{DecreasingRainfall_gaussian}-\autoref{Decreasing_AmNat};  and for sufficiently low precipitation these curves abruptly shift to a state with no vegetation. This abrupt transition comes much earlier for kernel $\rho_{1}$ that has higher width. Furthermore, the curves for simulations with cut-off Gaussian with higher width $\sigma$, coalesces with the curve for $\rho_{2}$. \par We have also carried out numerical simulations similar to those in \autoref{DecreasingRainfall_gaussian}-\autoref{Decreasing_AmNat} but with a higher rate of precipitation decay $(\frac{da}{dt}=-10^{-2},\frac{da}{dt}=-10^{-3})$ to better understand how the system-response is dependent on the rate of variation of precipitation level. Like in the earlier cases, here too as the precipitation decays, the system shifts from an initial vegetated state to patterned states and then to a complete desert state after a number of pattern rearrangements and wavelength adaptations. However, the system is found to be undergoing wavelength adaptations with increasing step-size for growing rates of variation in $a$; i.e., when the rate of change is higher, a much higher proportion of patches become extinct while undergoing wavelength adaptation compared to the cases in \autoref{DecreasingRainfall_gaussian}-\autoref{Decreasing_AmNat}. Due to this, desertification occurs at precipitation levels where stable patterned states previously existed at earlier instances. This infers the possibility of the rate-dependent transition, but a detailed study of this phenomenon is out of the scope of this work.  
\par
\begin{figure*}
	\centering
	
	\begin{minipage}{0.48\linewidth}
		\centering
		\includegraphics[width=\linewidth]{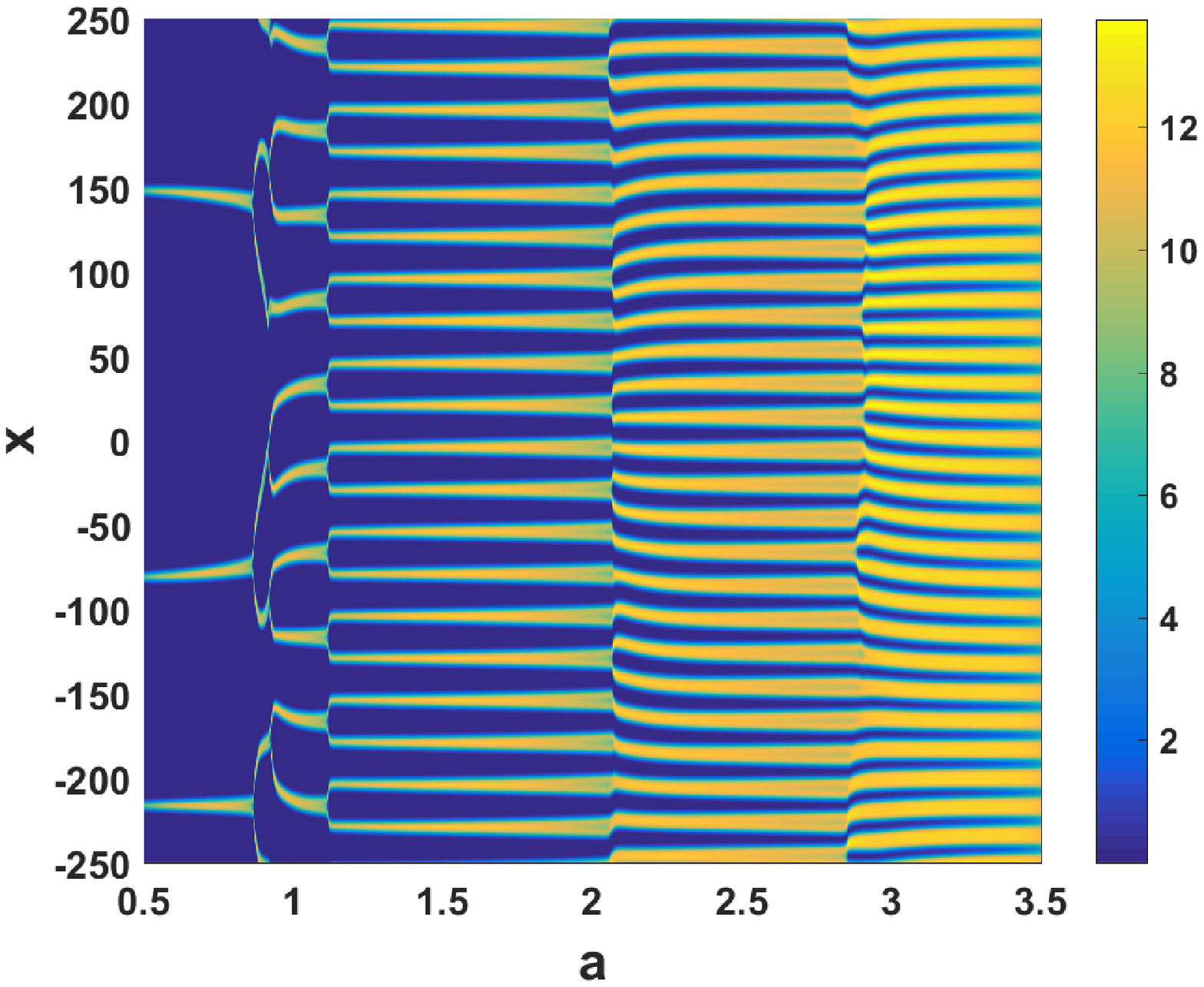}
		\subcaption{} 
		\label{restoration_sigma_1_type_II}
	\end{minipage}
	\begin{minipage}{0.48\linewidth}
		\centering
		\includegraphics[width=\linewidth]{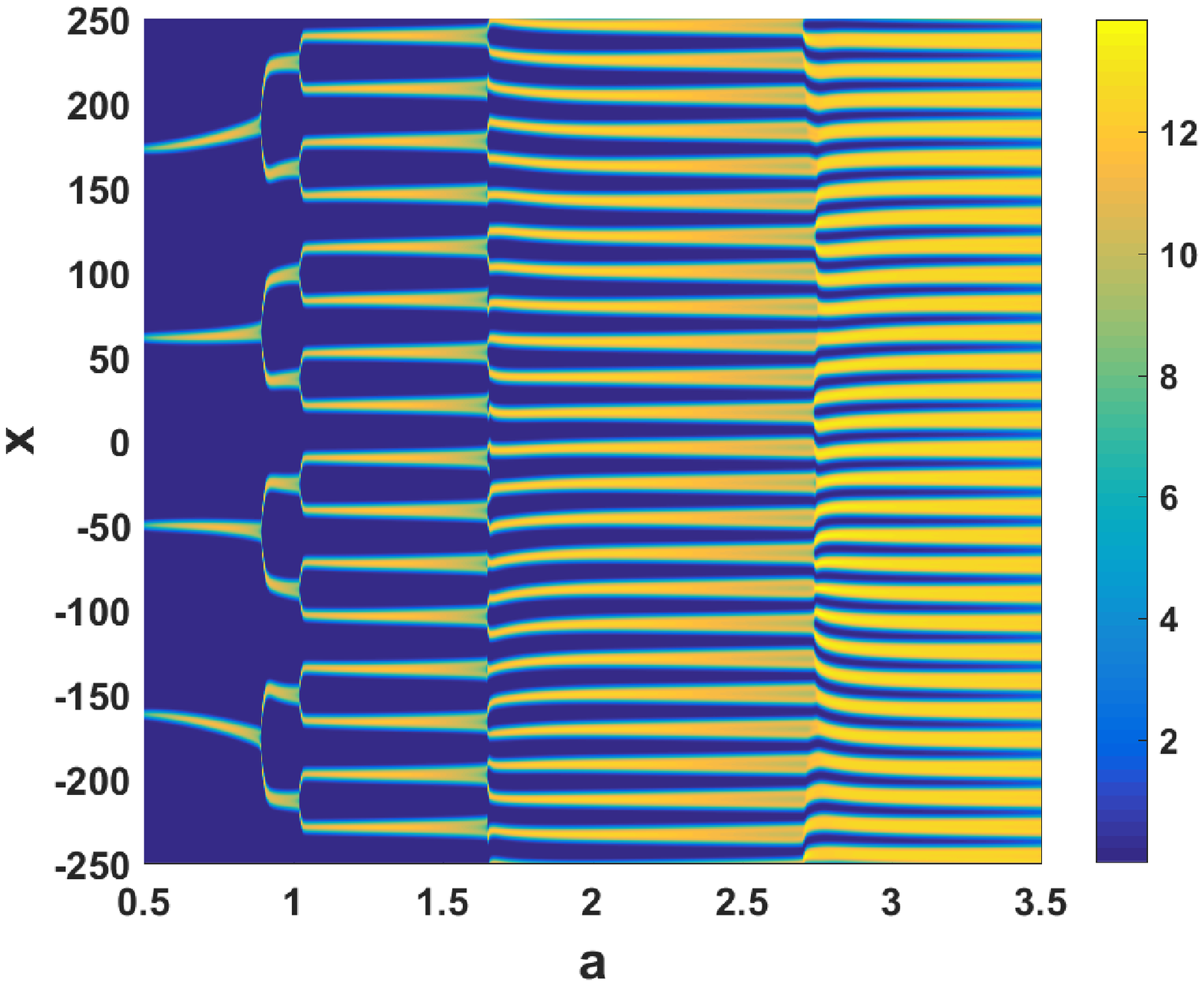} 
		\subcaption{}
		\label{restoration_sigma_1_type_III}
	\end{minipage}
	\newline
	\begin{minipage}{0.48\linewidth}
		\centering
		\includegraphics[width=\linewidth]{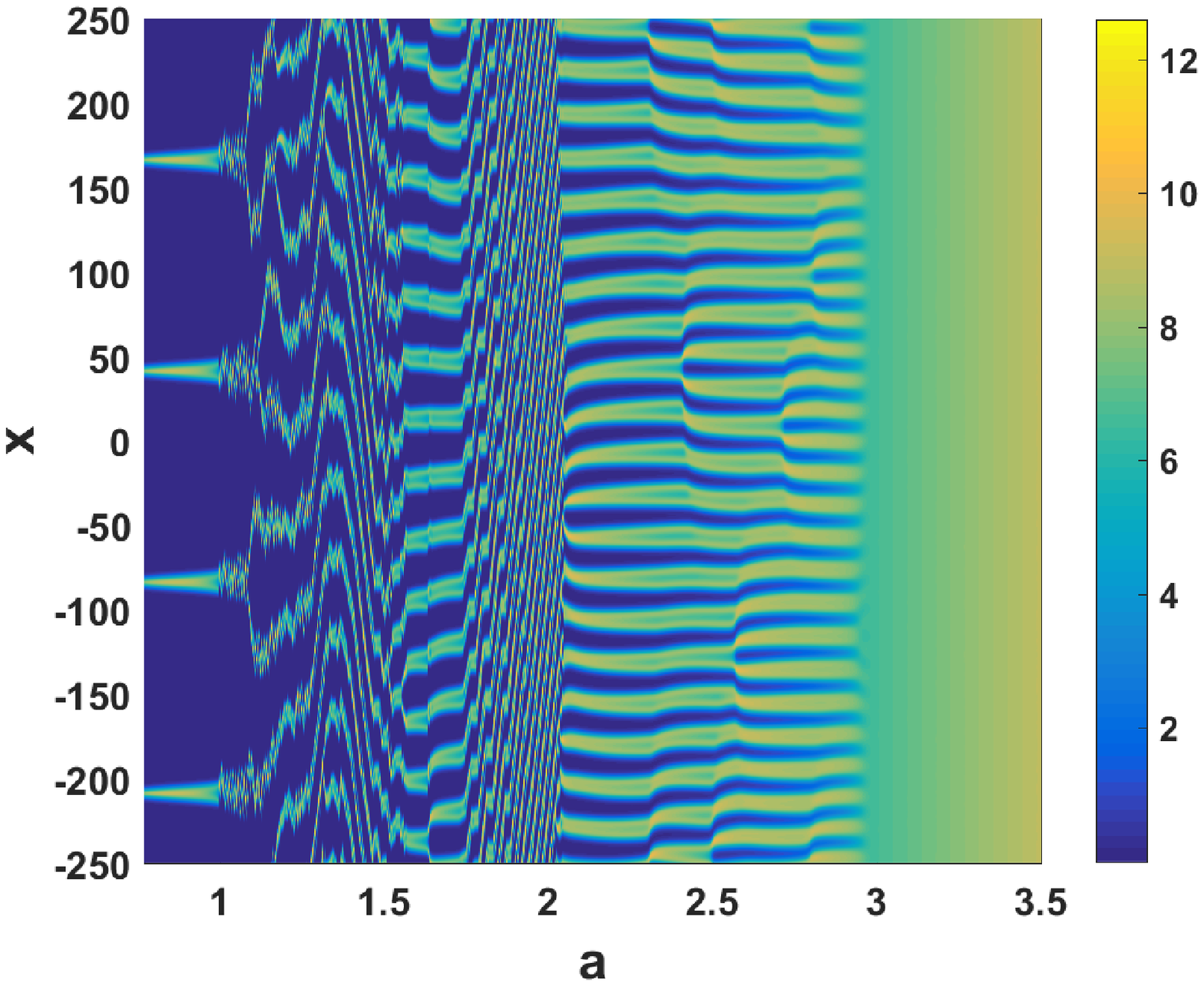}
		\subcaption{} 
		\label{restoration_sigma_8_type_II}
	\end{minipage}
	\begin{minipage}{0.48\linewidth}
		\centering
		\includegraphics[width=\linewidth]{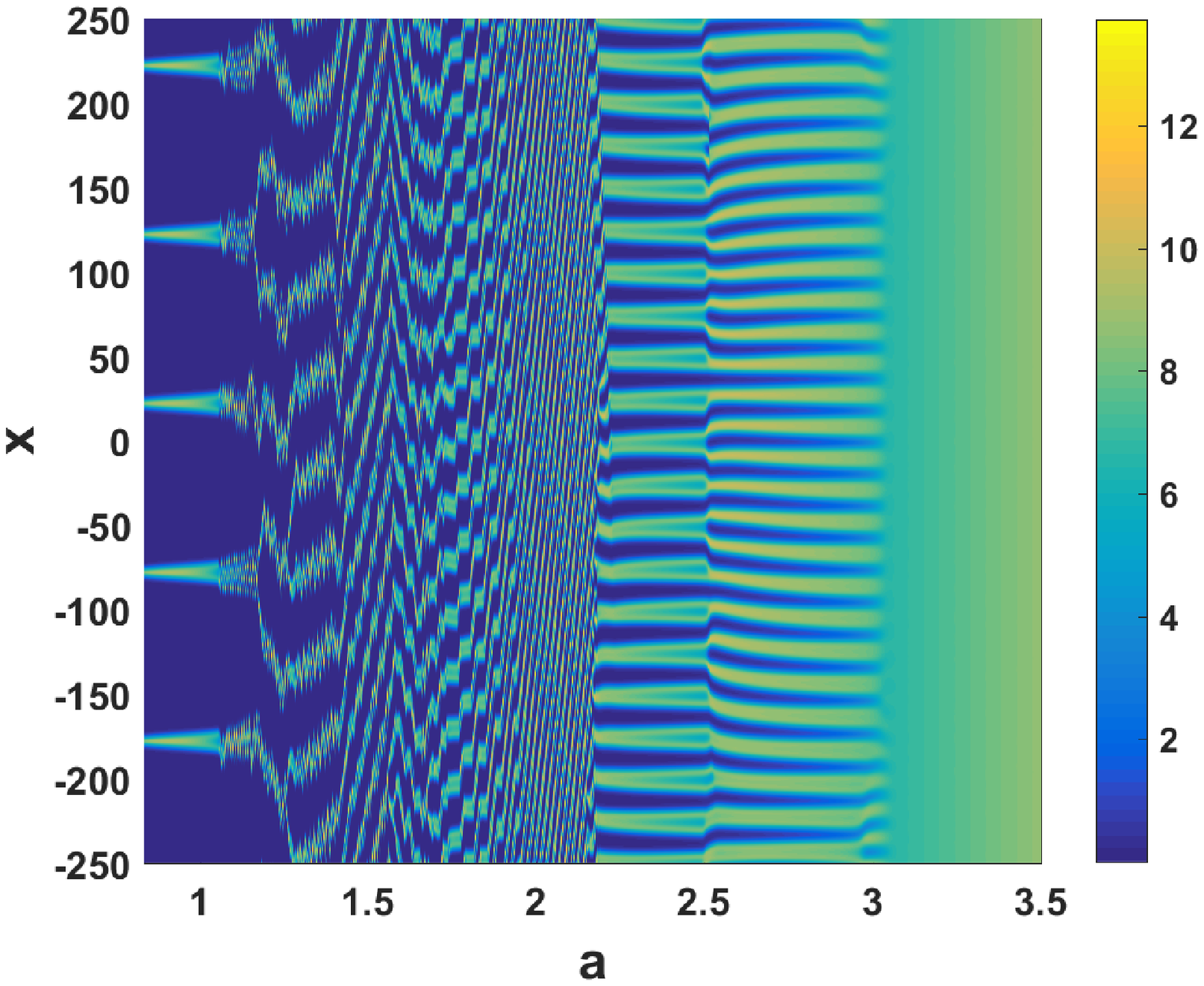} 
		\subcaption{}
		\label{restoration_sigma_8_type_III}
	\end{minipage}
	\newline
	\begin{minipage}{0.48\linewidth}
		\centering
		\includegraphics[width=\linewidth]{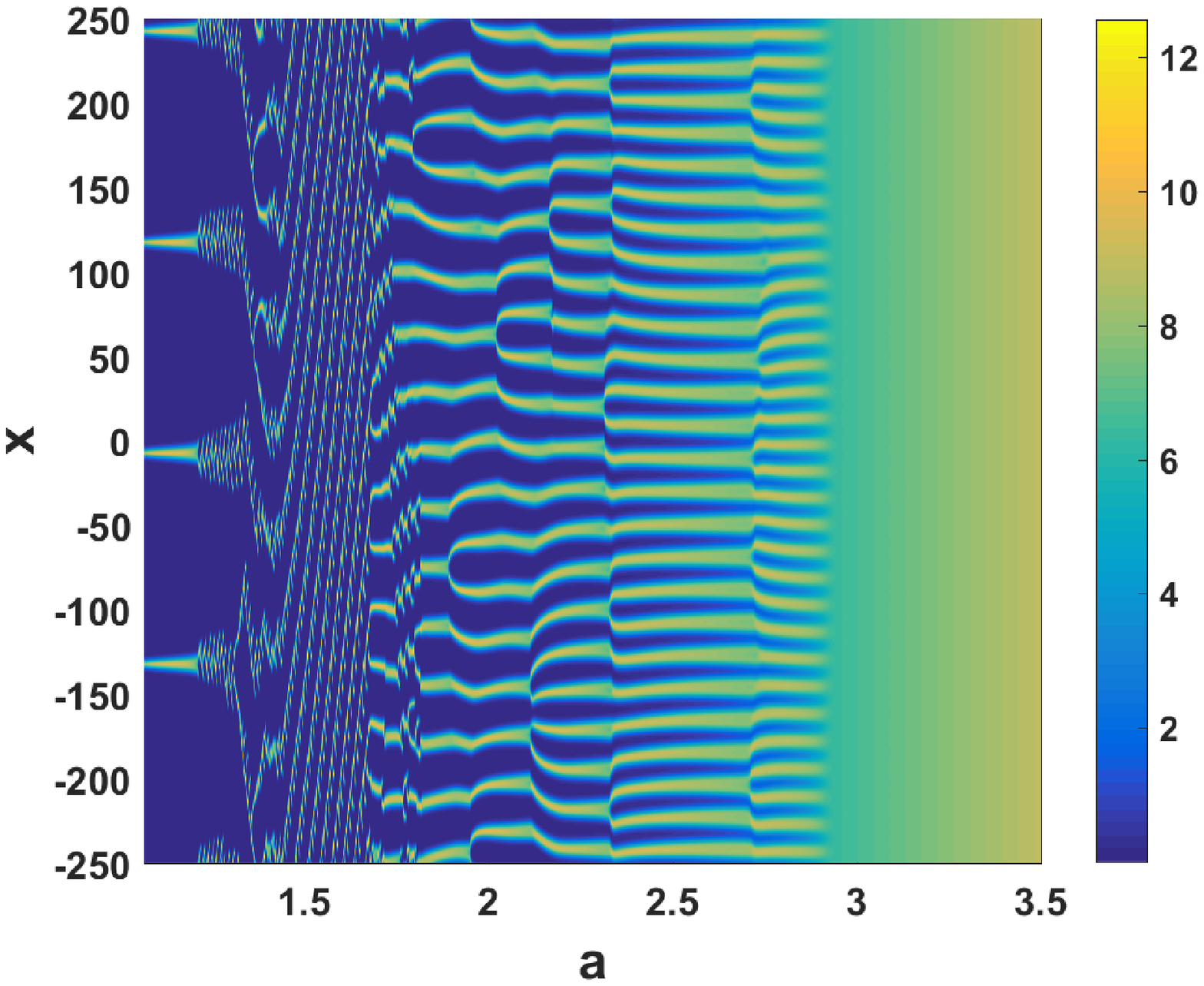}
		\subcaption{} 
		\label{restoration_sigma_15_type_II}
	\end{minipage}
	\begin{minipage}{0.48\linewidth}
		\centering
		\includegraphics[width=\linewidth]{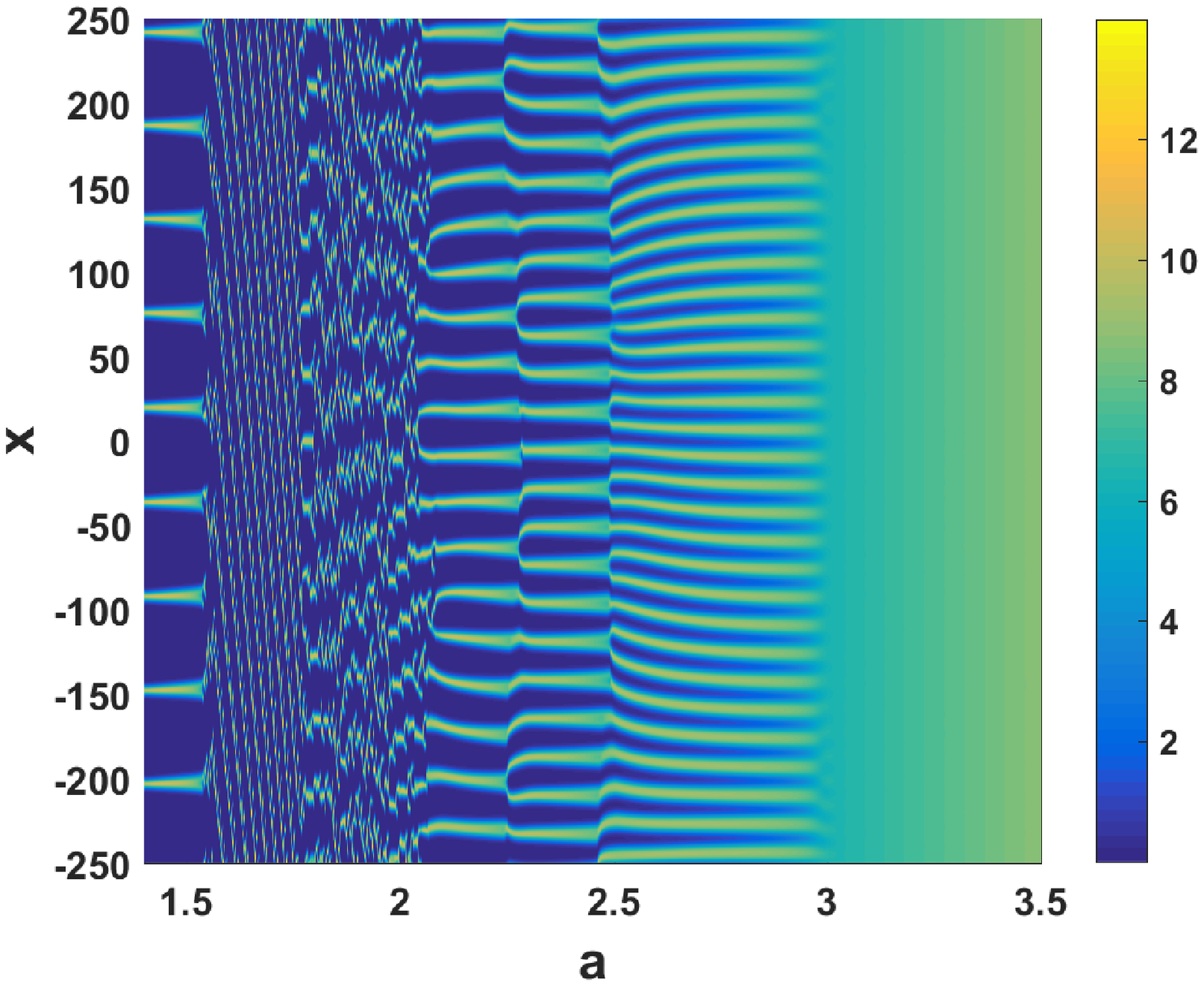} 
		\subcaption{}
		\label{restoration_sigma_15_type_III}
	\end{minipage}
	
	\caption{Simulation results for system (\ref{final}-\ref{KL_graz_fun}) with kernel $\rho_{1}$ (\ref{cut-off Gaussian}) having different $\sigma$ values: (a-b) $\sigma=1$, (c-d) $\sigma=8$, (e-f) $\sigma=15$. The left side column is for Sustained Grazing (a-c-e) and the second column (b-d-f) is for Natural Grazing. Initial vegetation distribution is taken from \autoref{DecreasingRainfall_gaussian} at precipitation where very few vegetation patches are left and then the precipitation level is set to increase at a rate $\frac{da}{dt}= 10^{-4}$}
	
	\label{IncreasingRainfall_gaussian}
\end{figure*} 
Earlier studies \cite{siero2019grazing,sherratt2013history} have demonstrated that extended Klausmeier model (\ref{KL_modified}) manifests hysteresis phenomena: numerical run for decreasing precipitation yielded patterns with shorter wavelength than patterns for increasing precipitation. To know the dependence of system-response on its history, simulations have been carried out with the same setting of \autoref{DecreasingRainfall_gaussian}-\autoref{Decreasing_AmNat}, but this time with precipitation increasing at a rate $\frac{da}{dt}=10^{-4}$. Following previous studies \cite{siero2019grazing,siteur2014beyond}, simulation is performed up to a precipitation level $a=3.5$. Vegetation distributions of \autoref{DecreasingRainfall_gaussian}-\autoref{Decreasing_AmNat}, where a very little amount of vegetation is left (i.e. at the precipitation level just before vegetation density $n$ becomes lesser than $10^{-6}$), are taken as the initial conditions. The simulation results (\autoref{IncreasingRainfall_gaussian}) for kernel $\rho_{1}$ having different width, show that with increasing precipitation, the initial vegetation distribution continues for some time and number of vegetated patches remains the same. Further increment in precipitation results in a sequence of transitions to patterns with shorter and shorter wavelength, and ultimately a regime shift to a state with homogeneous vegetation takes place. In case of $\sigma=1$, the system is unable to recover fully and no homogeneously vegetated state appears till $a=3.5$ for both type of grazing. But as the width $\sigma$ increases, the final transition from patterned state to uniformly vegetated state happens (\autoref{restoration_sigma_8_type_II}-\autoref{restoration_sigma_15_type_III}). This regime shift occurs for slightly higher level of precipitation for natural grazing, which reciprocates the earlier observations for decreasing precipitation. Moreover, this final regime shift occurs at precipitation level which is greater than the Turing bifurcation threshold $a_T$ in \autoref{DecreasingRainfall_gaussian}. Furthermore, as expected the simulation results for uniform kernel $\rho_{2}$ are in line with Ref.  \cite{siero2019grazing}; not shown here for the sake of brevity. It is observed that, unlike Ref. \cite{siero2019grazing}, the restoration of vegetation under increasing precipitation can be seen not only for natural grazing, but in case of sustained grazing also. Another noteworthy fact is, the wavelength of these patterns at any given precipitation level is larger with respect to their counterpart in \autoref{DecreasingRainfall_gaussian}. This can be understood simply by counting the number of ridges in \autoref{IncreasingRainfall_gaussian} at any particular value of $a$ and comparing them with their corresponding result in \autoref{DecreasingRainfall_gaussian}. So at any specific precipitation level, there are multiple possible stable states at which the system may reside, depending on its history.           
\section{Discussion and Conclusion}\label{Discussion}
This study aims to evaluate the influence of the intrinsic spatial non-locality in herbivore grazing over the response of dry-land ecosystem to changing environmental conditions. While modeling vegetation dynamics in arid ecosystem, Ref. \cite{siero2018nonlocal,siero2019grazing} have elegantly incorporated a mean-density dependent grazing response; here it is further modified by using a Gaussian weight function for deriving the mean vegetation density. This modification is driven by the simple assumption that a grazer at any location will be more influenced by the forage nearby than that of further away. It has been observed that such a simple ecological aspect of herbivory effects the dynamics of ecosystem significantly. \par 
Results of this study clearly reconfirm that apart from the water and nutrient availability, herbivory is also a major player in the  functioning of dry-land ecosystem. Simulations in \autoref{Response To Changing in Precipitation Level} show that the system responds to the change in precipitation by regulating vegetation biomass through self-organization of patterns. With increasing aridity, system starting from a homogeneously vegetated state ($\mathcal{V}$) first shifts to a patterned state due to Turing bifurcation and then undergoes a sequence of pattern adaptations to patterns with larger and larger wavelength, followed by a regime shift to completely barren desert state. In non-spatial models, with declining precipitation the system was thought to stay in stable uniformly vegetated state until a tipping point (marked by $*$ in \autoref{Fig1}) is reached, where a critical transition to the alternate stable state ($\mathcal{B}$) happens \cite{rietkerk2021evasion}. However, findings of this study suggest that the steady state ($\mathcal{V}$) with uniform vegetation cover, loses stability against spatial effects through Turing bifurcation, at much higher precipitation level than that of the tipping point and
vegetation persists through rearrangement of patterns for precipitation level beyond the tipping point, which reiterates the significance of self-organized patterns for maintaining productivity in dry-land ecosystem \cite{siteur2014beyond,bastiaansen2020effect}. One of the new phenomena observed in this study is that the system response to increasing environmental pressure heavily depends on the characteristic of grazer, for example perceptual abilities (e.g. sight, olfaction). From \autoref{Turing_prediction_region} it can be noticed that when the width $\sigma$ of the weight function ($\rho$) is small (i.e. scenario where grazer relies mainly on the vegetation present in a close vicinity), the shift from homogeneous vegetation cover to patterned state happens at relatively higher precipitation level. Moreover, for both sustained and natural grazing, complete desertification comes at more arid condition for the case with smaller width. From comparison of \autoref{DecreasingRainfall_gaussian} and \autoref{Decreasing_AmNat} it is evident that the ultimate critical transition to barren desert state takes place at more dry condition in case of Gaussian kernel, i.e. the grazing ecosystem is more resilient to increasing aridity than it was considered to be in previous studies \cite{siero2018nonlocal,siero2019grazing}. Another interesting fact observed in this study is that the decline in average vegetation density in response to the increasing aridity is not gradual, rather sudden oscillations can be observed in \autoref{Spatial_Avg} whenever the patterns rearrange themselves by adapting wavelength. Observations in Ref. \cite{siero2019grazing} suggest that in case of human controlled grazing, the final regime shift occurs at a high precipitation level (\autoref{AmNat_2}); however a very small amount of vegetation continues to exist for far more arid situation in case of natural grazing (\autoref{AmNat_3}). In contrast, this study exhibits resilience to decreasing precipitation in both type of grazing (\autoref{DecreasingRainfall_gaussian}). Another noteworthy fact is that as width of the Gaussian influence function increases, the dynamical behavior of the system become increasingly alike to the system with uniform kernel $\rho_{2}$ (which resembles the model in Ref. \cite{siero2018nonlocal,siero2019grazing}). This can be easily explained by comparing the shape of weight functions $\rho_{1}$ and $\rho_{2}$. With growing $\sigma$ the bell shape of Gaussian function gets more and more flattened, hence the weights all over the range becomes almost equal, which resembles the case for uniform kernel.\par Our findings also imply that the response of patterned ecosystems to environmental variation depends not only on the magnitude of the variation but also on the rate at which conditions change. Similar type of results had been observed in Ref. \cite{siteur2014beyond}. When the rate of change in precipitation level is rapid, the adaptation process is less gradual. Such circumstances prevent vegetation patches from rearranging and the system shifts abruptly to a completely degraded state for less arid conditions compared to the scenarios with a slow rate of change. So to tackle the increasing environmental pressure efficiently, it is also necessary to identify the critical rates of change in environmental conditions. \par 
This work demonstrates possibility of restoration in grazing ecosystem, which is in line with previous studies \cite{siteur2014beyond,sherratt2013history}. Model runs with increasing precipitation, exhibit complete recovery to uniformly vegetated state at higher precipitation level, preceded by a number of pattern transitions to patterns with shorter and shorter wavelengths. Unlike Ref. \cite{siero2019grazing}, improving environmental condition results in restoration of vegetation not only for natural grazing, but also for sustained grazing case. Furthermore, qualitative comparison between \autoref{DecreasingRainfall_gaussian} and \autoref{IncreasingRainfall_gaussian} reveals that although the rate of change in precipitation are same, the system response to increasing and decreasing precipitation are contrasting. This infers to the history-dependence of system response under stressful environment (better known as Hysteresis) and possibility of multi-stability in the grazing ecosystem, which reciprocates the observations of earlier studies \cite{siero2019grazing,siteur2014beyond}. \par 
Summarizing, this study reveals how the incorporation of herbivore-grazing as function of distance from the grazer, can substantially alter the ecosystem response to changing environmental condition. Apart from providing novel insights into grazing ecosystem, this study also reconfirms the findings of recent studies in arid ecosystem. Moreover, outcomes of this work necessitates inclusion of spatial non-locality while modeling herbivory in vegetation systems. However, it must be mentioned that this study is based on a phenomenological approach, intended to integrate the scale-dependent feedback and spatially non-local herbivore grazing. It can further provide theoretical framework for future data acquisition on herbivore foraging using sophisticated techniques and satellite images.
Assimilating the biological behavior-driven movement of herbivores with realistic data support while choosing weight function would be a research problem worth pursuing. Furthermore, in order to understand the transitions among qualitatively different patterns in the presence of herbivore grazing, the two dimensional version of this model needs to be explored. 
	
\begin{acknowledgements}
	This work is supported by UGC, Govt. of India under NET-JRF Scheme. We
	are grateful to Jasashwi Mandal and Swarnendu Banerjee for providing technical support. We would also like to thank anonymous reviewers for their insightful suggestions. 
\end{acknowledgements}	
	\bibliography{Vegetation_PRE_Revision}
	\bibliographystyle{apsrev4-1}

\end{document}